\documentclass[ 
 prx,
 amsmath,
 amssymb,
 reprint,
 superscriptaddress, 
 longbibliography,
]{revtex4-2}
\usepackage{graphicx}
\usepackage{dcolumn}

\usepackage[utf8]{inputenc}
\usepackage[T1]{fontenc}
\usepackage{mathptmx}
\usepackage{etoolbox}

\usepackage{bibentry}

\usepackage{siunitx}
\usepackage{float}
\usepackage{xfrac}
\usepackage{hyperref}
\usepackage{xcolor}
\xdefinecolor{tugreen}{RGB}{128, 186, 38}

\newcommand\sref[2]{\hyperref[#1]{\ref*{#1}(#2)}}

\usepackage{microtype}

\usepackage{placeins}

\usepackage{pdfpages} 
\usepackage{pgffor} 
\usepackage{xr}

\def\supplementfilename{SI}

\newif\ifarXiv
\arXivtrue

\makeatletter
\AtBeginDocument{\let\LS@rot\@undefined}
\def\@email#1#2{%
 \endgroup
 \patchcmd{\titleblock@produce}
  {\frontmatter@RRAPformat}
  {\frontmatter@RRAPformat{\produce@RRAP{*#1\href{mailto:#2}{#2}}}\frontmatter@RRAPformat}
  {}{}
}%
\makeatother

\begin{document}

\title{Temporal sorting of optical multi-wave-mixing processes 
in semiconductor quantum dots}


\author{S.~Grisard}
\email{stefan.grisard@tu-dortmund.de}
\affiliation{Experimentelle Physik 2, Technische Universit\"at Dortmund, 44221 Dortmund, Germany}

\author{A.~V.~Trifonov}
\affiliation{Experimentelle Physik 2, Technische Universit\"at Dortmund, 44221 Dortmund, Germany}

\author{H.~Rose}
\affiliation{Paderborn University, Department of Physics \& Institute for Photonic Quantum Systems (PhoQS), 33098 Paderborn, Germany}

\author{R.~Reichhardt}
\affiliation{Experimentelle Physik 2, Technische Universit\"at Dortmund, 44221 Dortmund, Germany}

\author{M.~Reichelt}
\affiliation{Paderborn University, Department of Physics \& Institute for Photonic Quantum Systems (PhoQS), 33098 Paderborn, Germany}

\author{C.~Schneider}
\affiliation{Technische Physik, Universit\"at W\"urzburg, 97074 W\"urzburg, Germany}
\affiliation{Institute of Physics, University of Oldenburg, 26129 Oldenburg, Germany}

\author{M.~Kamp}
\affiliation{Technische Physik, Universit\"at W\"urzburg, 97074 W\"urzburg, Germany}

\author{S.~H\"ofling}
\affiliation{Technische Physik, Universit\"at W\"urzburg, 97074 W\"urzburg, Germany}

\author{M.~Bayer}
\affiliation{Experimentelle Physik 2, Technische Universit\"at Dortmund, 44221 Dortmund, Germany}

\author{T.~Meier}
\affiliation{Paderborn University, Department of Physics \& Institute for Photonic Quantum Systems (PhoQS), 33098 Paderborn, Germany}
\author{I.~A.~Akimov}
\affiliation{Experimentelle Physik 2, Technische Universit\"at Dortmund, 44221 Dortmund, Germany}

\date{\today}

\begin{abstract}
Coherent control of ensembles of light emitters by means of multi-wave mixing processes is key for the realization 
of high capacity optical quantum memories and information processing devices. In this context, semiconductor quantum dots 
placed in optical microcavities represent excellent candidates to explore strong light-matter interactions beyond the limits 
of perturbative non-linear optics and control the unitary evolution of optically driven quantum systems. In this work, 
we demonstrate that a sequence of two optical picosecond pulses can be used to establish coherent control over the phase 
evolution of the ensemble of charged excitons (trions) in (In,Ga)As quantum dots independent of their initial quantum state. 
Our approach is based on coherent transfer between degenerate multi-wave-mixing signals in the strong field limit where 
Rabi rotations in multi-level systems take place. In particular, we use the two-pulse photon echo sequence to uncover the 
coherent dynamics of the trion ensemble, whereas the areas of two additional control pulses serve as tuning knobs 
for adjusting the magnitude and timing of the coherent emission. Furthermore, we make use of the spin degeneracy of ground 
and excited state of charged quantum dots to control the polarization state of the emitted signal. 
Surprisingly, we reveal that the use of optical control pulses, whose durations are comparable to the dephasing time of the ensemble, 
lifts the temporal degeneracy between wave-mixing processes of different order. This phenomenon is manifested in a significant modification 
of the temporal shape of the coherent optical response for strong optical fields, which is in accordance with the developed theoretical model. 
Lifting the temporal degeneracy allows to smoothly trace the transition from the perturbative to the regime of Rabi rotations
and opens up new possibilities for the optical investigation of complex energy level structures in so far unexplored material systems. 
\end{abstract}    

\maketitle
\section{Introduction}
Non-linear coherent optical spectroscopy and in particular multi-wave-mixing processes are of great interest 
for fundamental research in material science and applications in photonics~\cite{boyd_2020}. In this respect, nowadays special 
attention is drawn to applications in the field of quantum optics where non-linear optical phenomena allow to 
establish non-classical interferometry~\cite{chekhova_nonlinear_2016}, signal processing~\cite{willner_all-optical_2014}, 
and quantum optical 
memories~\cite{lvovsky_optical_2009, tittel_photon-echo_2010}. 
Under weak excitation, the efficiency of multi-wave-mixing signals strongly decreases with increasing number of interactions 
between the medium and the involved optical fields, i.e., the order of the underlying nonlinear optical effect. 
However, the situation changes drastically under 
resonant excitation with ultrashort optical pulses 
where multi-wave-mixing signals can be significantly enhanced~\cite{chemla_many-body_2001}. 
Prominent examples are multidimensional coherent spectroscopy and transient 
four-wave-mixing techniques that were successfully applied for investigations of the energy structure and relaxation dynamics 
of electronic excitations in solid state systems and in particular 
semiconductors~\cite{shah_2010, cundiff_optical_2013, kasprzak_coherent_2011, cundiff_optical_2009, hao_direct_2016}.
Here, the primary electronic excitation is represented by excitons with strong 
oscillator strength and pronounced many-body interactions which leads to a significant increase of high-order
non-linearities~\cite{yakovlev_exciton_2018, poltavtsev_photon_2018, katsch_optical_2020}.

Recent studies in semiconductor quantum dots (QDs) exploited multi-wave-mixing processes to evaluate the energy structure 
of exciton complexes that is otherwise hidden by strong inhomogeneous broadening. In particular, five-wave mixing from 
biexcitons has been demonstrated in the anti-Stokes coherent 
response~\cite{moody_fifth-order_2013} and six-wave mixing has been used to gain access to non-damped quantum beats in the 
exciton-biexciton system~\cite{tahara_generation_2014}. Except advances in higher harmonic generation~\cite{tritschler_evidence_2003} 
or studies of extrem nonlinear phenomena~\cite{tritschler_extreme_2003, mucke_role_2002}, most of the optical studies including atomic 
systems were however performed in the perturbative regime~\cite{zuo_generalized_2006, zhang_controlling_2007, zhang_coexistence_2008}. Of 
particular interest are degenerate multi-wave-mixing processes where the exciting optical fields have the same photon energy as the 
emitted signal. In this case for resonant excitation with intensive pulses, i.e. when Rabi rotations take place, there is always a 
set of wave-mixing signals  that are emitted simultaneously. In this work we reveal that this temporal degeneracy is partly lifted 
for a particular choice of the duration of the optical pulses which are used to generate the non-linear response. In this respect, it becomes 
possible to decouple different wave-mixing contributions and to smoothly trace 
the transition from perturbative to strong field regime with Rabi rotations 
by temporally resolving the optical response of the system under study.

Non-linear optical processes play a crucial role for coherently controlling the quantum state of the system 
and changing its optical response, which is appealing for the implementation of logical gates and ultrafast optical signal processing 
~\cite{stievater_rabi_2001, kamada_exciton_2001, suzuki_coherent_2016}. Therefore, an understanding of multi-wave 
mixing in the non-perturbative regime is of major importance. 
While most of the concepts are partly adopted from nuclear magnetic resonance 
(NMR) techniques, there are significant differences in the speed of control, polarization selectivity, and phase matching condition 
dictated by spin and photon momentum conservation~\cite{slichter_2011, smallwood_multidimensional_2018}. Using polarization degrees 
of freedom in the optical domain gives access to new and unique possibilities of control that are not available in NMR protocols. 
Semiconductor QDs represent an ideal playground for coherent control experiments in the optical domain~\cite{bayer_bridging_2019}. 
They allow to perform initialization and switching of quantum states at ultrafast sub-ps time 
scales~\cite{bonadeo_coherent_1998, flissikowski_two-photon_2004, henzler_femtosecond_2021} with 
well-defined polarization selection rules for optical transitions~\cite{dyakonov_2017} 
and robust coherence under resonant excitation with intensive optical pulses for establishing a strongly non-linear 
regime~\cite{zrenner_coherent_2002, ramsay_damping_2010, suzuki_coherent_2016, wigger_rabi_2018, grisard_multiple_2022}. Recently, 
switching between four- and six-wave mixing was demonstrated to control 
the state of a single QD~\cite{fras_multi-wave_2016}. Alternative approaches using pre-pulses that set the initial population 
of the excited states in a multi-level system demonstrated further possibilities for selective excitation 
and read-out using two-dimensional Fourier spectroscopy~\cite{suzuki_coherent_2016, suzuki_detuning_2018}. However, coherent 
control over the phase evolution of excitons and the role of detuning with respect to the resonant optical field in the multi-wave 
mixing processes remained unexplored. 

In the present work, we focus on optically controlling the phase evolution of an ensemble of singly charged (In,Ga)As QDs which is
manifested by the interplay of temporally sorted high-order multi-wave-mixing signals in a photon echo experiment. We introduce 
a sequence of two resonant optical pulses to manipulate the phase evolution of charged excitons (trions), whereas the 
independent two-pulse photon 
echo protocol serves as a tool to detect the phase evolution of the trion ensemble. Depending on the area of the control pulses, 
two main effects are revealed. First, two 
$\pi$-pulses invert the phase of each trion which leads to a transfer of the ensemble's coherent emission, i.e., the photon 
echo pulse, to a controllable moment of time. Second, two $\pi / 2$-pulses evoke a response of the ensemble that corresponds to the 
formation of Ramsey fringes in the spectral domain. In both cases, the signal is sensitive to the mutual optical phase between the 
control pulses which we incorporate here to manipulate the polarization state of the collective emission of trions that each consist 
of two spin degenerate ground and excited states. To that end, we demonstrate a splitting of the photon echo pulse in two crossed 
polarized pulses, which is accomplished by independent control of the phase evolution in each of the polarization paths. 
Furthermore, by temporally resolving the optical response from the trion ensemble, we 
reveal a comprehensive shape of the resulting photon echoes, which is
caused by the finite duration of the optical picosecond pulses. To explain this behavior we apply a theoretical modeling where we 
disentangle the nonlinear optical response into perturbative multi-wave-mixing contributions. Surprisingly, we find that
different multi-wave-mixing orders appear sorted at different moments of time allowing us to trace the transition in the 
formation of coherent optical response from the perturbative regime to the limit of strong field excitation.

\section{Temporal sorting of multi-wave-mixing processes from a trion ensemble}
\begin{figure}
    \centering
    \includegraphics[width = \columnwidth]{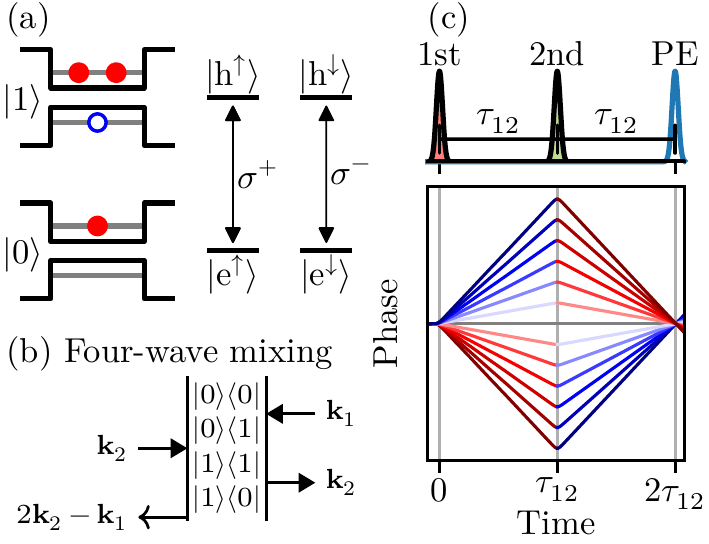}
    \caption{(a) Schematic illustration of ground state $|0\rangle$ and excited state $|1\rangle$ 
    in a charged quantum dot. The ground state consists of a resident electron in the conduction band (red), 
    the excited state is given by a trion state formed by two electrons (red) and one hole in the valence band (blue).
    Taking into account the spin degeneracy of the ground and excited state (electron/hole spin up or down, respectively), the system 
    can be treated as a four level scheme as shown on the right side, where the transitions between the two independent two level systems 
    can be excited by circular polarizations $\sigma^+$, $\sigma^-$. 
    (b) Exemplary double sided Feynman diagram representing the four-wave mixing response resulting from the interaction of a 
    two-level system with two laser fields with respective wavevectors $\mathbf{k}_i$. (c) Illustration of the 
    photon echo protocol, where the macroscopical polarization is rebuilt by inverting the phase evolution of individual emitters 
    as shown in the lower panel.}
    \label{fig: fig01}
\end{figure}  
In the following, we describe the approach that we propose for the observation of multi-wave-mixing 
signals arising from the interaction between four resonant laser pulses 
and a quantum dot ensemble. The basis of our approach is set by the two-pulse photon 
echo (2PE) technique that we 
expand by two additional optical pulses.  
In this way, we gain a high degree of control over the  
phase evolution of the ensemble.
Here, we make use of the unique property of inhomogeneously broadened ensembles that 
allow to separate different multi-wave-mixing components in time by appropriate choice of temporal 
delays between the applied pulses. Furthermore, we consider the spin degeneracy of ground and excited states 
in charged quantum dots to control the polarization state of different wave-mixing components. 
For the discussion of the general properties of the considered mechanisms, 
we analytically solve the optical Bloch equations assuming delta-like optical pulses 
(impulsive limit, subsection~\ref{sec: impulsive_limit}). 
Thereafter (subsection~\ref{sec: finite_pulses}), we discuss the effect of a finite pulse duration on 
the temporal shape of the resulting photon echo response. 
The latter becomes crucial for distinguishing between multi-wave-mixing contributions which appear 
at the same temporal position in case of infinitely short control delta-pulses.

The optical response of the experimentally studied 
quantum dots mainly arises from negatively charged quantum dots, 
i.e., trions~\cite{grisard_multiple_2022}. 
As schematically shown in Fig.~\sref{fig: fig01}{a}, the ground state is 
given by a single electron in the conduction band, 
whereas the excited state consists of two electrons with opposite spin projections in the conduction band 
and one hole in the valence band. 
Taking into account the two possible spin orientations of the electron in the ground state
and the hole in the excited state, 
the system can be represented by a four-level system (FLS) as shown in Fig.~\sref{fig: fig01}{a} with ground states 
$|\mathrm{e}^{\uparrow\downarrow}\rangle$ and excited states $|\mathrm{h}^{\uparrow\downarrow}\rangle$. 
The scheme is made up of two separated two-level systems (TLS) that can be independently addressed by the two circular 
light polarizations $\sigma^+$ and $\sigma^-$.
Under excitation with linearly co-polarized pulses and in absence of a magnetic field, the FLS behaves 
effectively like a TLS consisting of states $|0\rangle$ and $|1\rangle$~\cite{langer_magnetic-field_2012}. 
When, however, the relative amplitude or optical phase between $\sigma^+$ and $\sigma^-$
components of the exciting light is varied, the spin degeneracy 
of the trion in the ground and excited state offers possibilities for the polarization control of 
the coherent optical response of the quantum dot ensemble, as we will discuss below.

As a starting point, we briefly review the 2PE protocol which results, in lowest order perturbation theory, 
from a four-wave-mixing (FWM) response proportional to $\sim E_1^* E_2^2$
with two resonant laser pulses $E_i$ with respective wavevectors $\mathbf{k}_i$ that are temporally separated 
by a delay of $\tau_{12}$. 
The characteristics of the signal are represented by means of a double 
sided Feynman diagram in Fig.~\sref{fig: fig01}{b}. 
This type of diagram tracks the evolution of the elements of the density operator 
$\hat{\rho}_{ij} =  |i\rangle\langle j|$ ($i, j \in \{0, 1\}$) upon interaction with the involved light fields. 
During the FWM process, the first pulse creates a polarization $|0\rangle\langle 1|$ that freely evolves until the second pulse 
converts the polarization to a population $|1\rangle\langle 1|$ of the upper level. The population is again instantaneously converted 
to a polarization $|1\rangle\langle 0|$ by the same pulse that finally leads to a field emitted
in the phase-matched direction $\mathbf{k}_S = 2\mathbf{k}_2 - \mathbf{k}_1$. 
Regarding an inhomogeneous ensemble, the decay of the macroscopical 
polarization after action of the first pulse results from the fact that the involved TLS 
acquire a relative phase depending on their detuning $\nu$ from the laser frequency.
The motion of the individual phases is represented in Fig.~\sref{fig: fig01}{c}.
At time $\tau_{12}$, the second pulse inverts the phase of each individual oscillator. Therefore, the whole dephasing 
process runs backwards and a macroscopical polarization is rebuilt in form of a photon echo 
at time $2\tau_{12}$~\cite{hahn_spin_1950, kopvillem1963luminous, kurnit_observation_1964}. 

\subsection{Impulsive limit}\label{sec: impulsive_limit}
\begin{figure*}
    \centering
    \includegraphics[scale = 1]{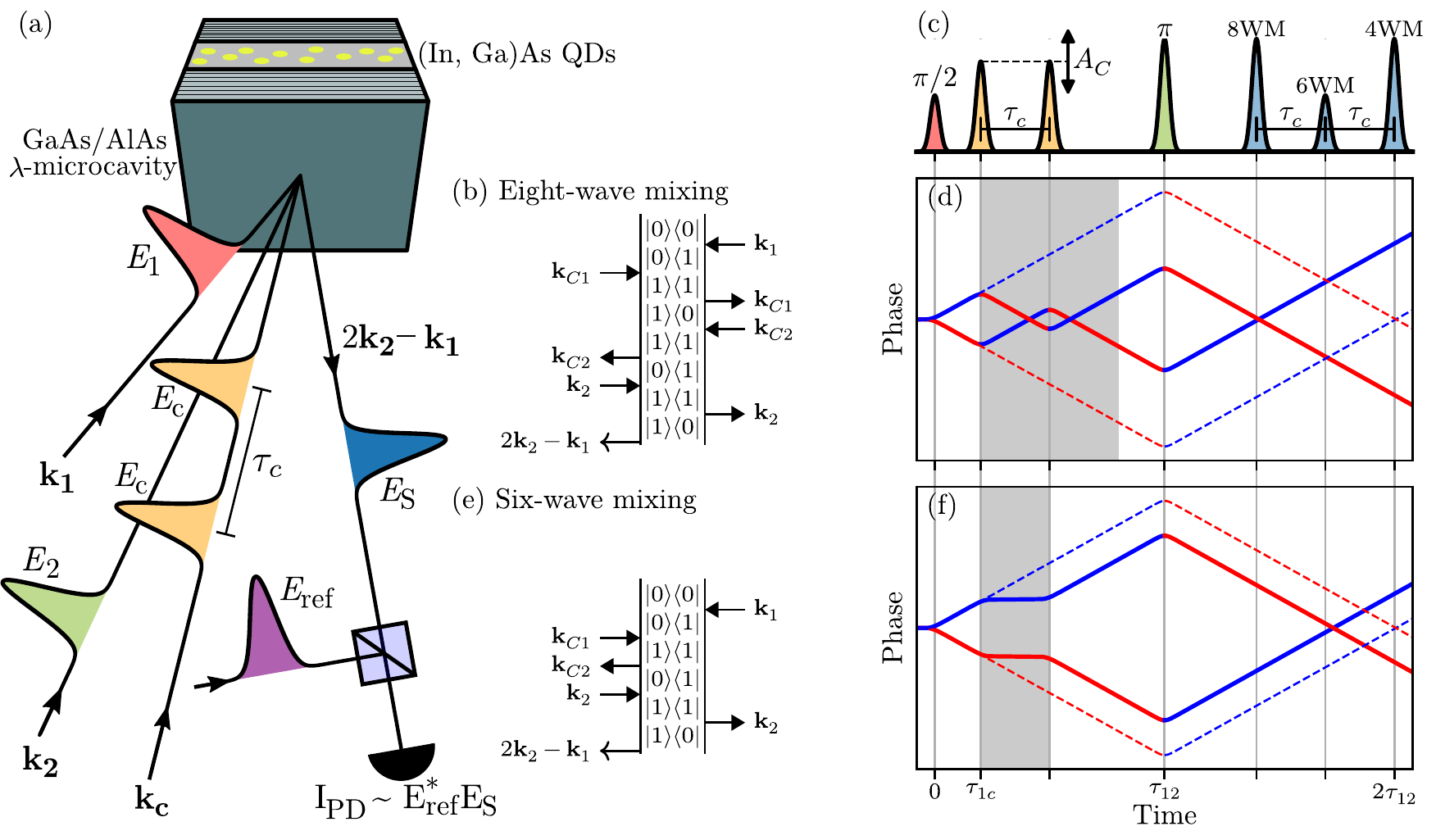}
    \caption{(a) Temporal and angular arrangement of the optical pulses for our experiments. The basis is set by the
    2PE experiment, where two pulses with wavevectors $\mathbf{k}_1$ and $\mathbf{k}_2$ 
    generate a photon echo signal in the 
    phase matched direction $2\mathbf{k}_2 - \mathbf{k}_1$. In between these two pulses we introduce two control pulses 
    sharing the same wavevector
    $\mathbf{k}_C$. The resulting coherent responses are temporally resolved through interference with a reference beam at 
    time $\tau_\mathrm{ref}$. The measured value is the current on a photo-diode, which is 
    proportional to $E_SE^*_\mathrm{ref}$. Note that all pulses possess the same photon energy and colors in the 
    figure are chosen only for clarity.
    (b)/(e) Double sided Feynman diagrams 
    for the eight- and six-wave-mixing processes as described in the main text.
    (c) Temporal arrangement of optical pulses for reference on the time axis of the phase diagrams in (d) and~(f). 
    (d)/(f) Evolution of the phase of two individual TLS for the two different wave-mixing processes in (b) and (e). Dashed lines show the 
    phase evolution in the absence of the control pulses for comparison.}
    \label{fig: fig02}
\end{figure*} 
We consider nonlinear responses such as the 2PE in the regime of coherent Rabi rotations, 
thus strongly exceeding the 
perturbative 4WM regime.
In this range, the 2PE amplitude depends in a non-monotonic manner on the areas 
$A_1$ and $A_2$ of the exciting and refocussing pulse, where the pulse area is defined as 
$A^\pm = \int \Omega_R^\pm(t) \mathrm{d}t$.
Here, $\Omega_R^\pm$ denotes the Rabi frequency $\Omega_R = \mu E_0^\pm / \hbar$, 
with the transition dipole moment $\mu$, the reduced Planck's constant
$\hbar$ and the slow varying envelope $E_0^\pm(t)$ of the 
time dependent electric field amplitude. With regard to the four level trion scheme, we 
introduced the superscript $\pm$ decomposing the electric field 
in $\sigma^+$ and $\sigma^-$ polarized components.
At first, we assume delta-like pulses, i.e. 
\begin{equation}
    E^\pm_i(t) = E^\pm_{0, i} \delta(t - t_i) \mathrm{e}^{i \phi^\pm},
    \label{eq: delta_pulses}
\end{equation}
and therefore neglect any dephasing, decoherence, and population decay during the action of each pulse.
In Eq.~\eqref{eq: delta_pulses}, 
we introduced the phase $\phi^\pm = \mathbf{k}\cdot\mathbf{r} + \varphi^\pm$ including the spatial phase $\mathbf{k}\cdot\mathbf{r}$
and an individual optical phase $\varphi^\pm$ for $\sigma^+$ and $\sigma^-$ components allowing, for example, 
to construct an arbitrary light polarization. 
Note that the finite duration of the pulses can only be safely neglected when the spectrum of the pulses 
is much broader than the inhomogeneous broadening of the ensemble, i.e. when the duration is much shorter than the reversible 
dephasing time $T_2^*$.
Although this condition is not fulfilled in our experiments, the consideration of the impulsive limit allows us to derive
compact analytical expressions for multi-wave-mixing signals whose general properties remain valid when the finite duration 
of the pulses is taken into account.
Therefore, we analytically solve the coupled equations of motion for the 
density matrix elements
$p^\pm = \langle \mathrm{e}^{\uparrow\downarrow} | \hat{\rho} | \mathrm{h}^{\uparrow\downarrow} \rangle$ and 
$n^\pm = \langle \mathrm{h}^{\uparrow \downarrow} | \hat{\rho} | \mathrm{h}^{\uparrow \downarrow} \rangle 
= 1 - \langle \mathrm{e}^{\uparrow \downarrow} 
| \hat{\rho} | \mathrm{e}^{\uparrow \downarrow} \rangle$, i.e. the Bloch equations in rotating wave approximation
\begin{subequations}
    \label{eq: OBE}
\begin{align}
    \frac{\mathrm{d}}{\mathrm{d}t} p^\pm &= i\frac{\left(\Omega_R^\pm\right)^*}{2}(1 - 2n^\pm) \mathrm{e}^{-i\phi^\pm - i \nu t} \\
    \frac{\mathrm{d}}{\mathrm{d}t} n^\pm & = \mathrm{Im} \left\{\Omega_R^\pm  p^\pm\mathrm{e}^{i\phi^\pm + i\nu t}\right\}
\end{align}    
\end{subequations}
Here, $\nu$ denotes the detuning, i.e. the difference between the frequency of the driving optical field and the 
frequency of the optical transition.
From a full solution for the optical coherences $p^+(t)$ and $p^-(t)$ being microscopic sources of $\sigma^+$ and 
$\sigma^-$ polarized light, 
we select only those contributions 
that contain the signal field wavevector $\mathbf{k}_S = 2\mathbf{k}_2 - \mathbf{k}_1$.
In this way, we arrive at the following expression of the 2PE fields in $\sigma^+$ and 
$\sigma^-$ polarization $E^\pm_\mathrm{2PE}$
\begin{equation}
    E^\pm_\mathrm{2PE} \sim \sin\left(A^\pm_1\right) \sin^2\left(\frac{A^\pm_2}{2}\right)  \mathrm{e}^{i(2\phi^\pm_2 - \phi^\pm_1) + i\nu(t - 2\tau_{12})}.
    \label{eq: FWM_Ai}
\end{equation}
A more detailed explanation of the calculation can be found in supplementary section~2. 
As can be seen from Eq.\eqref{eq: FWM_Ai}, the amplitude of the 2PE
oscillates as a function of both pulse areas, which is known as Rabi rotations~\cite{Allen.Eberly}. 
Due to the phase factor $\mathrm{e}^{i\nu(t - 2\tau_{12})}$, the averaging over a distribution of detunings $\nu$ in an 
inhomogeneous ensemble restricts the 2PE signal to the temporal position 
where all relative phases equal zero. The formation of macroscopical polarization of the ensemble can thus be observed at 
$t = 2\tau_{12}$ in form of a coherent light pulse. 

We aim to further optically manipulate the phase evolution of the quantum dot ensemble and in this way 
control the timing, amplitude, and polarization of the macroscopic optical response of the trion ensemble. 
For this purpose, we extend the 2PE protocol by two additional control pulses sharing the same wavevector 
$\mathbf{k}_{C1} = \mathbf{k}_{C2} = \mathbf{k}_C$ and 
electric field amplitude $E_C$. 
Fig.~\sref{fig: fig02}{a} sketches the temporal and angular arrangement of 
the, in total, four involved optical pulses. 
The control pulses are separated by a delay of $\tau_C$ and temporally located 
in between the exciting and refocusing pulse. 
The combination of four optical pulses results in 
a complex picture of photon echo responses corresponding to various multi-wave-mixing 
processes formed at different moments of time. For example, the first control pulse can act as refocusing pulse and create 
a photon echo in the direction $2\mathbf{k}_C - \mathbf{k}_1$.  
Here, we make use of the photon momentum conservation, to select 
only those responses that are emitted in the direction of the 2PE $2\mathbf{k}_2 - \mathbf{k}_1$, where the 
wavevector of the control pulses $\mathbf{k}_C \neq \mathbf{k}_2 \neq \mathbf{k}_1$ 
does not affect 
the overall phase matching condition. In this way, we only study those responses that are independent on the optical phase of the 
control pulses or only depend on the mutual phase between the control pulses. If necessary, other photon echo signals 
may be efficiently suppressed by choosing a large difference between $\mathbf{k}_C$ and $\mathbf{k}_1$/$\mathbf{k}_2$, 
such that the energy conservation is not fulfilled in the quantum dot cavity system with quadratic dispersion of the photonic mode. 

For a TLS, the contribution of each additional control pulse with wavevector $\mathbf{k}_{Ci}$ to the overall 
phasematching condition 
is $0, \, \pm 2\mathbf{k}_{Ci}$, or $\pm \mathbf{k}_{Ci}$. 
Therefore, the constraint on the phase matching condition $\mathbf{k}_S = 2\mathbf{k}_2 - \mathbf{k}_1$ not being affected by 
the control pulses may be realized in three
distinct scenarios, for which we explicitly write the phasematching conditions as 
\begin{subequations}
\begin{align}
    \mathbf{k}_S &= 2\mathbf{k}_2 - \mathbf{k}_1 + \mathbf{k}_{C1} - \mathbf{k}_{C1} + \mathbf{k}_{C2} - \mathbf{k}_{C2}, \label{eq: case1}\\
    \mathbf{k}_S &= 2\mathbf{k}_2 - \mathbf{k}_1 + 2\mathbf{k}_{C1} - 2\mathbf{k}_{C2}, \label{eq: case2}\\
    \mathbf{k}_S &= 2\mathbf{k}_2 - \mathbf{k}_1 + \mathbf{k}_{C1} - \mathbf{k}_{C2} \label{eq: case3}.
\end{align}    
\end{subequations}
Under the assumption $\mathbf{k}_{C1} = \mathbf{k}_{C2}$, all three cases fulfill the condition $\mathbf{k}_S = 2\mathbf{k}_2 - \mathbf{k}_1$.
In total, we therefore expect three distinct echo signals whose characteristics we discuss in the following. Each response leads to 
a macroscopic signal emitted at a well defined temporal position.   

The first case, Eq.~\eqref{eq: case1},
implies an insensitivity of the resulting signal to the optical phases of the control pulses 
and can be observed for any choice of $\mathbf{k}_{C1}$ and $\mathbf{k}_{C2}$ in the direction $2\mathbf{k}_2 -\mathbf{k}_1$. 
Within the impulsive limit, the temporal position of the resulting coherent emission will not be shifted relative to the 2PE
at $t =  2\tau_{12}$. 
As for the 2PE, we derive 
the dependence of the signal fields in $\sigma^\pm$ polarizations on the involved pulse areas 
\begin{equation}
    E^\pm_\mathrm{FWM} = E^\pm_\mathrm{2PE} \cos^4\left(\frac{A^\pm_C}{2}\right),
    \label{eq: EWM_Ai_2t12}
\end{equation}
where the functional dependence of $E^\pm_\mathrm{2PE}$ on $A_1$ and $A_2$ is given by Eq.~\eqref{eq: FWM_Ai}.
The signal is maximum for $A_C = 2 n \pi$ and zero for $A_C = (2n + 1) \pi$, where $n$ is an integer.  
We indexed the signal with FWM (four-wave mixing) corresponding to the lowest non-zero wave mixing order of the signal. We 
underline however, that~\eqref{eq: EWM_Ai_2t12} is the exact solution including all possible higher order wave-mixing 
components that share the 
same phasematching condition and temporal characteristic. 
As can be seen in Eq.~\eqref{eq: EWM_Ai_2t12}, the pulse area $A_C$ directly modulates
the amplitude of the 2PE, which may be also used to independently modify the contributions of 
$\sigma^+$ and $\sigma^-$ components as we will experimentally demonstrate in this work. 

Next, we consider the phasematching condition $\mathbf{k}_{S} = 2\mathbf{k}_2 - \mathbf{k}_1 + 2\mathbf{k}_{C1}
-  2\mathbf{k}_{C2}$, Eq.~\eqref{eq: case2}, 
which simplifies to the phase matching condition of the 2PE for the special case $\mathbf{k}_{C1} = \mathbf{k}_{C2}$.
This phasematching condition is realized, in lowest order perturbation theory, by an eight-wave mixing (EWM) process as 
depicted by an exemplary Feynman diagram in Fig.~\sref{fig: fig02}{b}.
Similar to the action of the refocussing pulse in the 2PE sequence, here, each of the 
control pulses converts the density matrix 
element to its complex conjugate, 
i.e. $|i\rangle\langle j|$ to $|j\rangle\langle i|$. 
The effect of this process on the overall coherent response is represented by the phase diagram in Fig.~\sref{fig: fig02}{d} in 
comparison to the 2PE (dashed lines). For a better overview, the diagram
plots the phase evolution of the field amplitude associated with 
only two oscillators with opposite detuning (blue and red).  
A coherent emission of the whole ensemble will appear only at the temporal positions where 
the relative phase of the oscillators disappears (crossing points of blue and red lines). 
For clarity, Fig.~\sref{fig: fig02}{c} on top 
of the phase diagram sketches the temporal arrangement of the pulses. 
The ensemble experiences a refocusing dynamic after the first control pulse 
that is again inverted by the second control pulse. In this way, the phases of individual oscillators are effectively 
unaffected for a time of $2\tau_c$ after the 
first control pulse, which we highlight by the grey area in Fig.~\sref{fig: fig02}{d}.   
Consequently, the refocusing pulse acting at $\tau_{12}$ generates a coherent emission that is shifted by 
$-2\tau_{c}$ with respect to the 2PE at $2\tau_{12}$.
Note that the two other moments where the relative phase is zero within the grey area in Fig.~\sref{fig: fig02}{d}
lead to a coherent emission in the directions $2\mathbf{k}_C - \mathbf{k}_1$ and $\mathbf{k}_1$, respectively, and are therefore not 
detected in our experiments. 
We again derive an analytical expression for the electric fields 
as a function of the involved pulse areas
\begin{equation}
    E^\pm_\mathrm{EWM} = E^\pm_\mathrm{2PE} \sin^4\left(\frac{A^\pm_C}{2}\right) \mathrm{e}^{2i \left(\phi^\pm_{C2} - \phi^\pm_{C1}\right) + 2 i\nu \tau_c}.
    \label{eq: 8WM}
\end{equation}
The dependence of the signal on $A_C$ 
oscillates with opposite phase as compared to
the signal at 
$2\tau_{12}$ $\left(E^\pm_\mathrm{FWM} \sim\cos^4\left(\sfrac{A^\pm_C}{2}\right)\right)$, 
Eq.~\eqref{eq: EWM_Ai_2t12}. 
Maxima appear for $A^\pm_C = (2n + 1)\pi$, minima for $A^\pm_C = 2n\pi$.
Consequently, 
the area $A_C$ acts as control knob for the coherent transfer between the two
echoes occurring at different temporal positions. For $A_C = \pi$, the control pulses function as a 
temporal gate for the coherent 
emission of the ensemble. Furthermore, the optical phase of the signal~\eqref{eq: 8WM} 
depends on twice the relative phases between the 
control pulses $2(\phi^\pm_{C2} - \phi^\pm_{C1})$, which allows, for example, to acquire full control over the 
polarization of the coherent emission by suitable choice of the polarizations of the 
control pulses as we experimentally demonstrate below. 

Lastly, we consider the phasematching condition $\mathbf{k}_{S} = 2\mathbf{k}_2 - \mathbf{k}_1 +\mathbf{k}_{C1}
-  \mathbf{k}_{C2} = 2\mathbf{k}_2 - \mathbf{k}_1$, Eq.~\eqref{eq: case3},
corresponding to a six wave mixing (SWM) response in lowest order perturbation theory. 
A corresponding Feynman diagram is 
presented in Fig.~\sref{fig: fig02}{e}. We again show the phase evolution of two oscillators 
with opposite detuning 
in Fig.~\sref{fig: fig02}{f}.
Here, the dephasing after the action of the first pulse is 
turned off within a duration of $\tau_C$ (grey area) as the control pulses convert 
polarization $|0\rangle\langle 1|$ to 
population $|1\rangle\langle 1|$ and back. 
Therefore, the coherent emission appears at $2\tau_{12} - \tau_C$. The functional dependence 
on the pulse areas is given by 
\begin{equation}
    E^\pm_\mathrm{SWM} = -\frac{E^\pm_\mathrm{2PE}}{2}\sin^2 \left(A^\pm_C\right) \mathrm{e}^{i \left(\phi^\pm_{C2} - \phi^\pm_{C1}\right) + i\nu \tau_c}, 
    \label{eq: 6WM}
\end{equation}
which runs through maxima for $A_C = \frac{(2n + 1)\pi}{2}$. 
Again, the optical phase of the signal can be controlled by the 
relative phase between the control pulses. 

\begin{figure}
    \centering
    \includegraphics[width = \columnwidth]{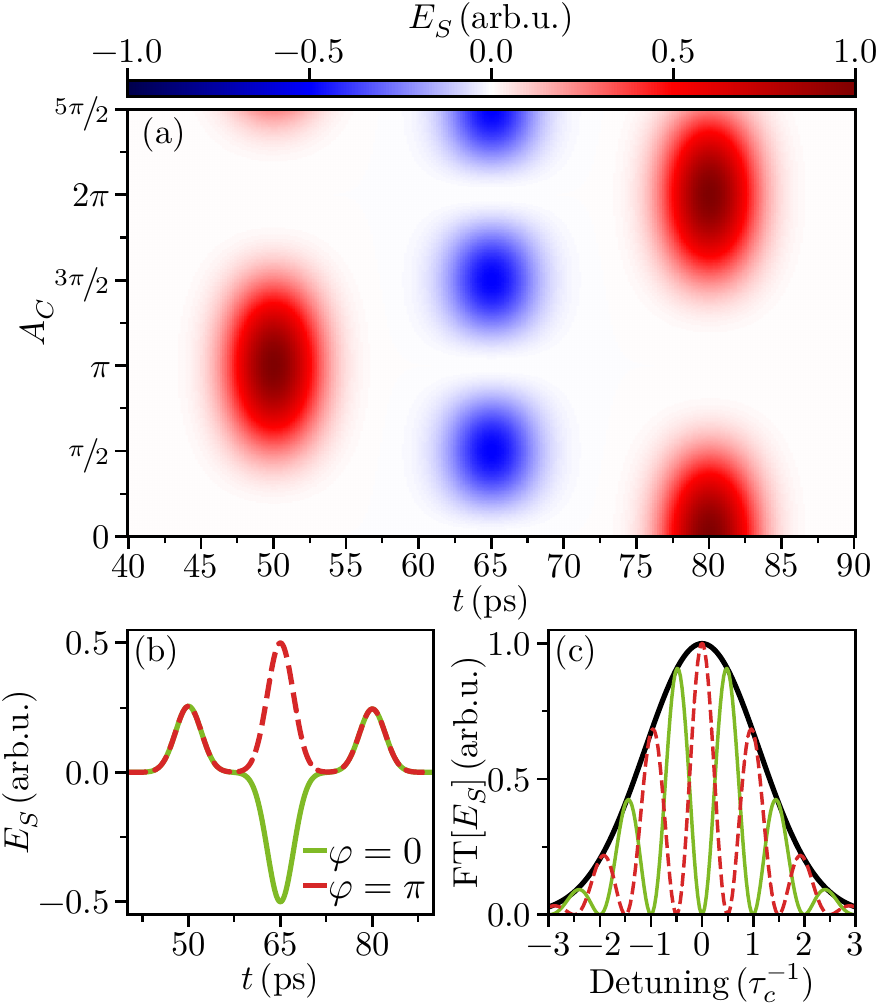}
    \caption{(a) Dependence of the signal field amplitude $E_S$ as a function of time $t$ and the area of the control 
    pulses $A_C$ in impulsive limit according to Eq.~\eqref{eq: totalsignal}. (b) 
    Temporal profile of the signal field $E_S$ for $A_C = \sfrac{\pi}{2}$
    and two different values of the relative phase $\phi$ between the two control pulses. 
    (c) Amplitude of the Fourier spectra of the temporal profiles shown in (b). 
    The black curve shows the distribution of detunings within the considered inhomogeneous ensemble of emitters. }
    \label{fig: fig03}
\end{figure}

In summary, we have introduced an arrangement of optical pulses 
that evoke different multi-wave-mixing contributions where two control pulses act as 
gates that can be used to control the amplitude, temporal position, and polarization of the coherent emission.
The overall signal fields $E^\pm_{S}$ in the phase matched direction $2\mathbf{k}_2 - \mathbf{k}_1$ can be written as 
the sum
\begin{equation}
    E^\pm_{S} = E^\pm_\mathrm{FWM} + E^\pm_\mathrm{SWM} + E^\pm_\mathrm{EWM}
    \label{eq: totalsignal}
\end{equation}
which is depicted in Fig.~\sref{fig: fig03}{a} by means of a two dimensional colormap as a function of real time 
and pulse area of the control pulses. Here, we assumed linearly co-polarized pulses, where the trion 
scheme works effectively as a TLS. The relative optical phase between the control pulses 
is set to zero. 
We further chose $\tau_{12} = \SI{40}{\pico\second}$, $\tau_{c} = \SI{15}{\pico\second}$, 
and a Gaussian inhomogeneous broadening of detunings $\nu$ of \SI{0.3}{\milli\eV}, equivalent to $T_2^* = \SI{4}{\pico\second}$ 
corresponding to the experimental situation 
(discussed below). 
For this choice, we can see well separated Gaussian photon echo signals at 
$2\tau_{12} = \SI{80}{\pico\second}$
($E_\mathrm{FWM}$), $2\tau_{12} - \tau_C = \SI{65}{\pico\second}$ ($E_\mathrm{SWM}$), 
and $2\tau_{12} - 2 \tau_C = \SI{50}{\pico\second}$ ($E_\mathrm{EWM}$) that oscillate as a function of $A_C$.

We would like to highlight the case $A_C = \sfrac{\pi}{2}$ that is in direct analogy to a Ramsey fringe experiment, which was 
studied as a demonstration of coherent control over the quantum state of a single quantum 
dot and an ensemble~\cite{stufler_ramsey_2006, michaelis_de_vasconcellos_coherent_2010, khanonkin_room-temperature_2021}.
In a standard Ramsey fringe experiment, a QD is exposed to two 
temporally separated $\pi / 2$ pulses with a detuning $\nu$ relative to the transition frequency of the quantum dot.  
When probing, for example, the population of the excited state as a function of the detuning or the 
temporal delay between the 
pulses, oscillations can be observed that are known as Ramsey fringes. 
The combination of two $\sfrac{\pi}{2}$ pulses with the 2PE sequence allows to observe 
Ramsey fringes from the inhomogeneous ensemble of emitters where different detunings 
are realized at once. Here, the Ramsey fringes manifest themselves in time domain in form of 
a temporal shape of the emission deviating from a single Gaussian pulse.   
We show temporal cross sections for $A_C = \sfrac{\pi}{2}$ of the response of the quantum dot ensemble 
in Fig.~\sref{fig: fig03}{b} for two different values of the relative phase between the control pulses 
$\phi = 0$ and $\phi = \pi$. The overall response consists of three Gaussian pulses 
centered at $2\tau_{12} - 2\tau_c$, $2\tau_{12} - \tau_c$, and $2\tau_{12}$.
As follows from Eq.~\eqref{eq: 6WM}, the sign of the 6WM response at 
$2\tau_{12} - \tau_c = \SI{65}{\pico\second}$ is inverted for a phase shift of $\phi = \pi$ with respect to the case $\phi = 0$. 
In the Fourier spectrum, Fig.~\sref{fig: fig03}{c}, these temporal cross sections correspond to Ramsey fringes with period 
$\tau_c^{-1}$ (here $(\SI{15}{\pico\second})^{-1} \approx \SI{67}{\mega\hertz}$) modulated by the Gaussian distribution of detunings. 
The phase of the fringes is directly set by the relative phase between the control pulses. 

\subsection{Effect of finite pulse durations}\label{sec: finite_pulses}
In this section we discuss how the results presented for the impulsive limit 
are modified when the pulse duration $t_p$ is not significantly smaller than the 
dephasing time $T_2^*$. In this case, the dephasing of the TLS during pulse 
action has to be taken into account. 
For our discussion, we consider an 
experiment with only one control pulse under otherwise the same conditions as in 
section~\ref{sec: impulsive_limit}.
In this way, we gain an intuitive understanding for the effect of finite pulse durations that can be 
expanded for more complex pulse arrangements. 

\begin{figure*}
    \centering
    \includegraphics[scale = 1]{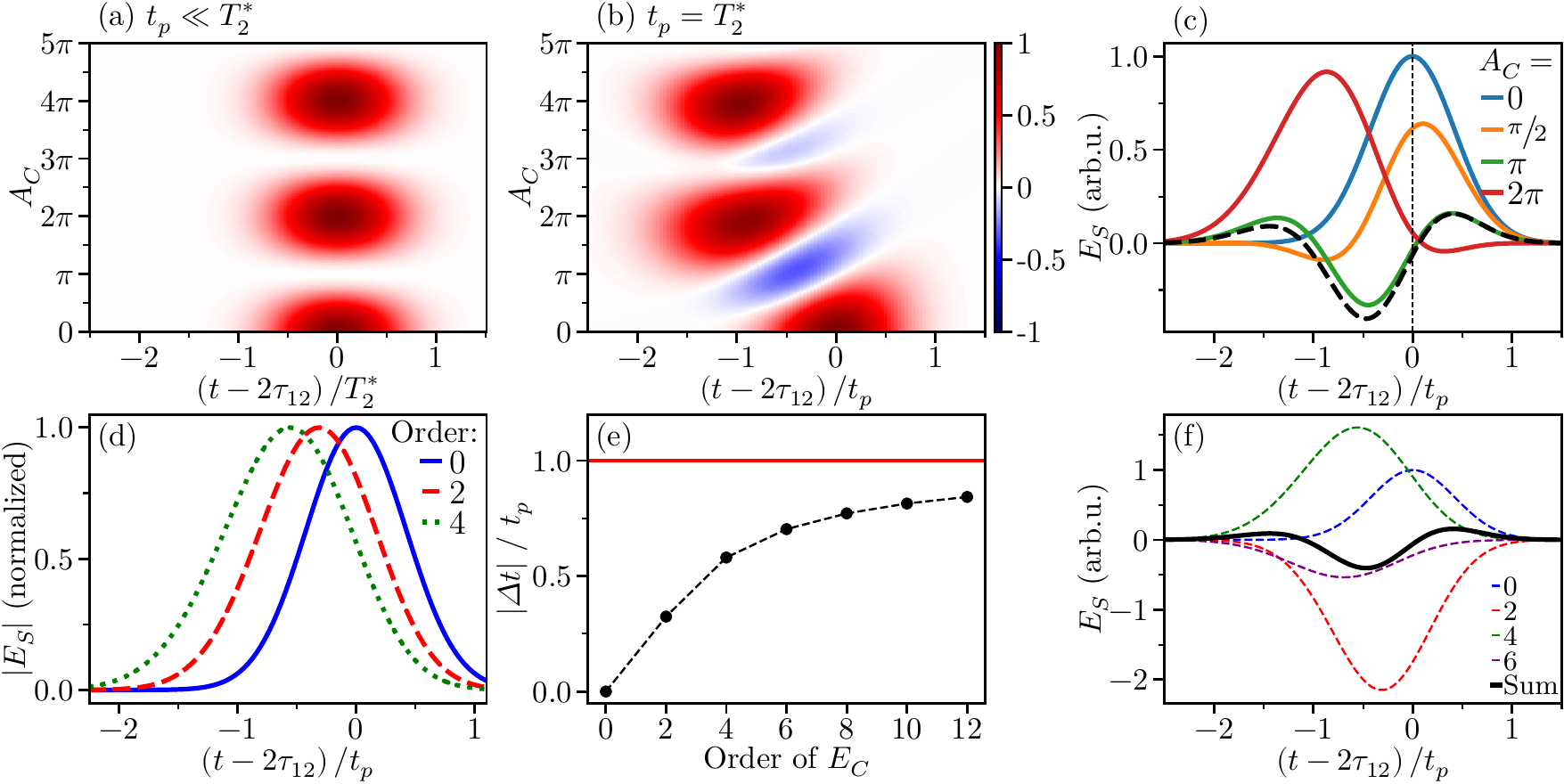}
    \caption{(a) Dependence of the signal amplitude $E_S$ as a function of time $t$ and the area of the single control 
    pulse $A_C$ in impulsive limit according to Eq.~\eqref{eq: single_control}. The time axis is shown relative to $2\tau_{12}$ in units 
    of the dephasing time $T_2^*$. (b) Calculation of $E_S$ for 
    a single control pulse with finite duration $t_p$ using optical Bloch equations. The time axis is shown in units of $t_p$. 
    (c) Temporal cross sections of the map shown in (b) for selected 
    values of the pulse area $A_C$. For comparison, the dashed black line shows the result 
    of a perturbative expansion of the signal field for the case $A_C = \pi$ (corresponds to the black line in (f)). 
    (d) Photon echo signals calculated for different numbers of interactions with the control pulse field. 
    (e) Temporal shifts of echoes calculated for different multi-wave-mixing orders. 
    (f) Decomposition of total signal for $A_C = \pi$ in up to ten-wave mixing components.}
    \label{fig: fig04}
\end{figure*}

We start again from an analytical expression for the full signal detected in the direction 
$2\mathbf{k}_2 - \mathbf{k}_1$ 
in the impulsive limit
\begin{equation}
    E_\mathrm{S} = E_\mathrm{2PE}  \cos^2\left(\frac{A_C}{2}\right),
    \label{eq: single_control}
\end{equation}  
which describes Rabi rotations of the echo signal 
as a function of the control pulse area at the fixed temporal position $2\tau_{12}$.
Equation~\eqref{eq: single_control} is visualized in 
Fig.~\sref{fig: fig04}{a} as a function 
of real time and pulse area $A_C$. Note that the time axis is normalized to the inhomogeneous dephasing time $T_2^*$. 

Next, we want to model the effect of a control pulse with finite duration and therefore choose 
$t_p = T_2^*$, i.e., a pulse duration equal to the duration of the photon echo which corresponds to the 
experimental situation shown later. The exciting and refocussing pulses 
are considered in the impulsive limit. For simplicity, the temporal shape of the control pulse is considered rectangular. For a 
rectangular pulse shape, an analytical  
solution of the optical Bloch equations is well 
known~\cite{Allen.Eberly}
and can be used to model the dependence of the coherent response 
as a function of detuning $\nu$, pulse area $A_C$ and real time $t$ in units of the pulse 
duration $t_p = T_2^*$. 
A closed expression for the signal field is given in Eq.~(S6) of the supplementary material. 
The final average over a Gaussian distribution of 
detunings with FWHM of $t_p^{-1}$ is carried out numerically. 
The result is graphically presented in Fig.~\sref{fig: fig04}{b} and can be directly compared to the case $t_p \ll T_2^*$ in~(a). 
As in the impulsive limit, we can observe Rabi rotations of the signal with maxima 
at $A_C = 2n\pi$. However, the maxima for 
$A_C = 2\pi,\,4\pi$ appear advanced by the pulse duration with 
respect to the 2PE for $A_C = 0$, as we highlight in Fig.~\sref{fig: fig04}{c}
where we plot temporal cross sections for selected values of $A_C$. 
For non-zero pulse areas below roughly $\sfrac{\pi}{2}$, the maximum of the coherent emission is 
shifted to positive delay times as shown in Fig.~\sref{fig: fig04}{c} by the orange curve corresponding to $A_C = \sfrac{\pi}{2}$. 
We can observe an additional pronounced negative
local extremum for $A_C = \pi$ where the signal equals zero in the impulsive limit, 
as given by Eq.~\eqref{eq: single_control}. 
The described advancement of photon echo was observed in Ref.~\cite{kosarev_accurate_2020} and can be explained 
by the fact that the action of a pulse with area of $2n\pi$ and $\Omega_R \gg \nu$ 
leaves the phases of individual TLS unaffected after a full Rabi rotation induced by the control 
pulse during $t_p$. Therefore, the overall phase evolution is effectively "frozen" for a duration of $t_p$ and 
the coherent emission of the ensemble is shifted. However, the temporal shifts for arbitrary pulse areas and in 
particular for $A_C \leq \pi/2$, where a retardation of the signal is observed, remained so far unclear.

The characteristic differences between Figures~\sref{fig: fig04}{a} and~\sref{fig: fig04}{b} can be understood by expanding the 
non-perturbative echo response in multi-wave-mixing orders and considering independently the effect of the finite pulse duration on their temporal behaviour.
For this purpose, we use the formalism developed by Mukamel et al.~\cite{mukamel_1999, schweigert_simulating_2008} that 
allows to calculate the $n$th order polarization $P^{(n)}$, being the source of a ($n$+1)-wave mixing response, using 
\begin{align}
    \begin{aligned}
    P^{(n)}(t) &\sim \int_{-\infty}^{t} \mathrm{d}t_n \int_{-\infty}^{t_n} \mathrm{d}t_{n - 1} \dots \int_{-\infty}^{t_2} \mathrm{d}t_{1} R^{(n)}(t, t_n, \dots, t_1) \\
     &\times \mathcal{E}_n(t_n)\dots \mathcal{E}_1(t_1),
    \end{aligned}
    \label{eq: P_n}
\end{align}    
where $\mathcal{E}_i(t_i) = \frac{1}{2} E_i(t_i) \mathrm{e}^{i\mathbf{k}_i \mathbf{r}} + \mathrm{c.c}$ are the temporal envelopes of the 
electric fields interacting with 
the system at time $t_i$. Here, also multiple interactions with the same field are allowed. 
$R^{(n)}$ is the nonlinear response function that is constructed from commutator products of the dipole operator $\hat{\mu}$
\begin{equation}
    R^{(n)} = \left(\frac{i}{\hbar}\right)^{\!\!n} \mathrm{Tr}\left([\dots[\hat{\mu}(t), \hat{\mu}(t_n)], \dots , \hat{\mu}(t_1)] \hat{\rho}\right),
    \label{eq: R_n}
\end{equation}
including summands that can be represented by double sided Feynman diagrams as shown above. 
A detailed analysis is presented in section 3 of the supplementary material. 

Using Eq.~\eqref{eq: P_n}, we calculate the nonlinear response in different orders with respect to the electric field amplitude of the control pulse and subsequently integrate over a Gaussian distribution of detunings $\nu$ with FWHM of $\sfrac{1}{t_p}$. The exciting and refocussing pulse are
considered as delta-pulses in lowest non-vanishing order, 
i.e. the first pulse in linear and the refocussing pulse in quadratic order. 
For the control pulse, we assume a rectangular shaped envelope of duration $t_p$. 
Note that we only consider the signal in the 
phase matched direction $2\mathbf{k}_2 - \mathbf{k}_1$, thus only those Feynman diagrams that include the same number of interactions with $E_C$ and $E_C^*$ contribute to the signal.
In Fig.~\sref{fig: fig04}{d} we compare the echo signals that involve zero, two, and four interactions with the control pulse, i.e. four-wave mixing ($E_S \sim E_1 E_2^{*2}$), six-wave mixing ($E_S \sim E_1 E_2^{*2} E_CE_C^*$), and eight-wave mixing ($E_S \sim E_1 E_2^{*2}E_C^2E_C^{*2}$) signals. 
Each perturbative order leads to the formation of a Gaussian shaped echo pulse. Interestingly, the temporal position is 
shifted towards negative delays in the case of six- and eight-wave mixing with respect to the four-wave mixing echo pulse 
at $t = 2\tau_{12}$.  
Note that we normalized the signals that were calculated for a fixed value of the 
electric field $E_C$ to their maximum. Due to the factor $(i)^n$ in definition~\eqref{eq: R_n}, the global sign of the 
wave-mixing signals alternates with increasing order. This effect was experimentally observed in Ref.~\cite{meier_signatures_2000} where 
different orders of wave mixing processes in pump probe experiments were compared. 

To further analyze the observed trend, 
we plot the temporal shift $\Delta t$ for up to twelve interactions 
(sixteen-wave-mixing) 
with the control pulse field in Fig.~\sref{fig: fig04}{e}. Here, $\left|\Delta t\right|$ increases monotonically  
tending towards the pulse duration $t_p$. Consequently, the temporal shift between two consecutive wave mixing orders 
approaches zero with increasing order.
We thus can interpret the complex waveforms of the coherent response of the ensemble with increasing 
area of the control pulse in Fig.~\sref{fig: fig04}{b}
as an interference pattern of wave-mixing orders that each give rise to a single photon 
echo pulse at a specific temporal position. 
Since the temporal shift is largest between consecutive low order wave-mixing components (Fig.~\sref{fig: fig04}{e}),
the interference pattern deviates most strongly from a single Gaussian echo pulse in the range of 
small pulse areas where low perturbative orders dominate the signal. 
As an example for the formation of non-Gaussian wave forms from the interplay of different wave-mixing orders, 
we plot in Fig.~\sref{fig: fig04}{f} the wave-mixing signals up 
to the sixth order in $E_C$ for $A_C = \frac{\mu}{\hbar} E_C  t_p = \pi$. 
The sum of the pulses with alternating sign results in the black solid curve that 
accurately reproduces the exact calculation as we highlight by directly comparing the exact and 
perturbative result (green and black line) in Fig.~\sref{fig: fig04}{c}.
The negative six-wave mixing signal (red dashed line in Fig.~\sref{fig: fig04}{f}) has the largest amplitude  
and is therefore dominantly responsible for the negative signal at $A_C = \pi$ that we mentioned above.

In summary, we have discussed a nonlinear optical regime, where the expansion of Rabi rotations into wave mixing orders allows 
to gain meaningful insights into the non-trivial temporal shape of coherent emission caused by the finite duration of optical 
pulses. In the impulsive limit, such expansion does not make sense since all wave mixing orders 
appear simultaneously and result in Rabi rotations centered around a fixed emission time. 
We found that the finite pulse duration however lifts this temporal degeneracy and leads 
to a temporal sorting of wave-mixing orders. For the experiments including two control pulses, we expect similar 
modifications of the impulsive limit, i.e. temporal shifts of the echo signals and the occurrence of new local extrema evoked 
by higher order multi-wave-mixing components. 

\FloatBarrier

\section{Demonstration of multi-wave mixing in co-polarized configuration}
\begin{figure*}
    \centering
    \includegraphics[scale = 1]{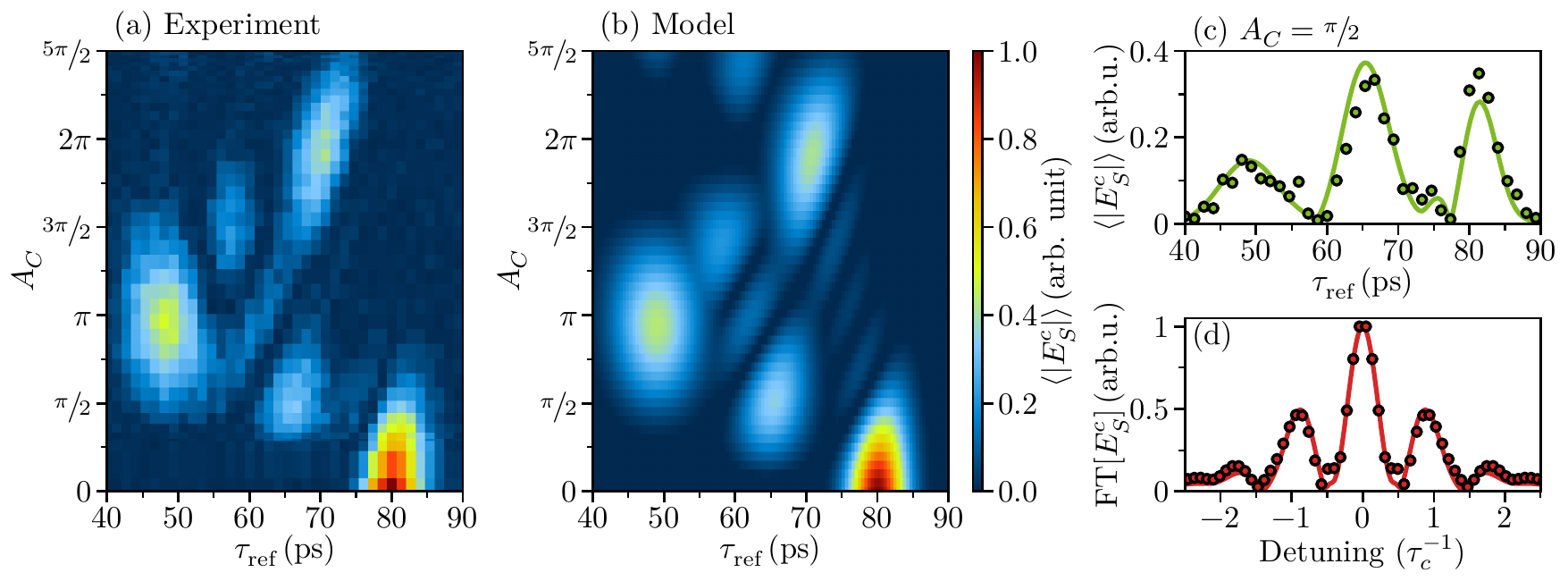}
    \caption{(a) Experimental dependence of the heterodyning signal on the area of the control pulses $A_C$ 
    and the reference time $\tau_\mathrm{ref}$. (b)
    Result of the calculation of $\langle|E_S^c|\rangle$ using optical Bloch equations 
    assuming rectangular shaped temporal envelopes of the control pulses. 
    The intensity-dependent damping mechanisms were considered phenomenologically as described in the main text. 
    (c) Temporal cross section of the color maps in (a) and (b) for $A_C = \sfrac{\pi}{2}$. Dots and lines corresponds to experiment and 
    calculation, respectively.
    (d) Absolute value of the Fourier transform of the temporal cross sections shown in (c).}
    \label{fig: fig05}
    \end{figure*}
Within this section, we neglect the polarization degree of freedom 
by choosing linearly co-polarized pulses. As mentioned above, the trion scheme can be considered as a TLS 
under this condition.   
In this way, we first set the focus on the temporal characteristics of the discussed multi-wave-mixing signals.

The experiments were performed on ensembles of (In,Ga)As/GaAs quantum dots (QDs) at a temperature of $\SI{2}{\kelvin}$.
The sample consists of a single layer of (In,Ga)As QDs 
embedded in an AlGaAs $\lambda$-microcavity.
The photonic mode of the microcavity is in resonance with the QD ensemble 
at $\SI{1.351}{\eV}$.
The photoluminescence spectrum of the QD-cavity system has a full width at half maximum (FWHM) of 
\SI{5.9}{\milli\eV}~\cite{grisard_multiple_2022}. 
The homogeneous linewidth of quantum dots amounts to roughly \num{1.6}\,{\textmu}eV, 
corresponding to a coherence time of $T_2 = \SI{0.83}{\nano\second}$~\cite{grisard_multiple_2022}.
We excite the sample resonantly using optical 
pulses that are generated by a titanium-sapphire laser with a repetition rate of \SI{75.75}{\mega\hertz} and have a duration of 
$t_p = \SI{4}{\pico\second}$ associated with the FWHM of the amplitude of their electric field
(corresponds to a duration of the intensity profile of \SI{2.8}{\pico\second}).    
The spectral width of the pulses, \SI{0.3}{\milli\eV}, is significantly narrower than the spectrum of the QD-cavity system but still much 
broader than the homogeneous linewidth of the quantum dots. Therefore, the laser pulses act on a macroscopic 
subensemble of emitters whose inhomogeneous broadening is 
governed by the laser spectrum. Note that this situation where the effective dephasing time $T_2^*$ 
equals the pulse duration is ideal for studying the effect 
of finite pulse durations on the coherent response that we discussed in section~\ref{sec: finite_pulses}.
The area of the laser pulses is adjusted by changing their intensity using 
combinations of $\lambda / 2$ retardation plates and polarizers. 
Since we study multi-wave-mixing signals as a function of the pulse area, it is decisive to address 
all quantum dots with the same intensity. 
Therefore, we use spatially flat intensity profiles of the control pulses as proposed in 
Ref.~\cite{grisard_multiple_2022}.
Under this condition we avoid a fading of the signal as a function of pulse area which is present for 
spatially Gaussian laser pulses~\cite{poltavtsev_damping_2017}. 

We temporally resolve the coherent response from the ensemble using the optical heterodyning technique where we capture the 
interference between the signal of interest $E_S$ and a reference pulse $E_\mathrm{ref}$ on a photo-diode, 
as shown in Fig.~\sref{fig: fig02}{a}.
Temporal profiles of $E_S$ are measured by scanning the 
delay $\tau_\mathrm{ref}$ of the reference pulse relative to the first pulse.
The measured signal on the photo-diode is proportional to the 
cross-correlation between $E_S$ and $E_\mathrm{ref}$, which we denote as 
$E_S^{c}(\tau_\mathrm{ref})$ in the following
\begin{equation}
    E_S^{c}(\tau_\mathrm{ref}) \sim \int E_\mathrm{ref}^*(t - \tau_\mathrm{ref}) E_S(t)  \, dt.
    \label{eq: CC}
\end{equation} 
By modulating the optical frequencies of the first and reference pulse using acousto-optical modulators, we isolate the 
phase-matched signal in frequency domain from noise and other wave-mixing signals and 
detect it using a lock-in amplifier. 
Further details on our experimental methods can be found in Ref.~\cite{poltavtsev_photon_2018}.

We introduced a temporal delay between exciting and refocussing pulse 
of $\tau_{12} = \SI{40}{\pico\second}$ and between exciting and the first control pulse of 
$\tau_{1c} = \SI{13}{\pico\second}$. The temporal gap between the control pulses is $\tau_c = \SI{15.3}{\pico\second}$ 
(temporal cross correlations between the reference and the control pulses are shown in section~1 of the Supplement). 
In this way, both control pulses interact with the sample
temporally in between the exciting and refocussing pulse and the temporal overlap between all pulses is negligible. 
Note that all temporal gaps are here significantly smaller than the coherence time of the quantum 
dots $T_2 = \SI{0.83}{\nano\second}$. The optical phases between all involved pulses are not actively stabilized. 
A phase-locked measurement of the signal fields is thus not subject to our studies. 
Instead, we capture the mean value of the modulus of the signal field $\langle|E_S^c|\rangle$, where $\langle \cdot \rangle$ 
denotes here the average over a uniform distribution of optical phases of the involved pulses. 

To investigate the influence of the control pulses on the temporal shape of the coherent response, 
we measured the heterodyning signal as a function of the reference time $\tau_\mathrm{ref}$ and the area of the control 
pulses $A_{C}$. For $A_C=0$ the coherent response is represented by the 2PE which is well described as Gaussian pulse 
centered at $\tau_\mathrm{ref} = 2\tau_{12} = \SI{80}{\pico\second}$. The control pulse area is varied by changing the 
intensity of the laser pulses. 
The areas of exciting and refocussing pulse were fixed at $A_1 = \sfrac{\pi}{2}$ and $A_2 = \pi$, respectively, 
corresponding to the maximum of the 2PE, Eq.~\eqref{eq: FWM_Ai}. The experimental result is presented as a two-dimensional 
color map in Fig.~\sref{fig: fig05}{a}. Next to the unshifted photon echo at $\SI{80}{\pico\second}$ for $A_C = 0$, we can observe 
four local maxima of the signal within the temporal range between $\SI{50}{\pico\second}$ and $\SI{80}{\pico\second}$ occurring at 
multiples of $A_C = \sfrac{\pi}{2}$.
This observation is in agreement with the modeling in the impulsive limit (see Fig.~\sref{fig: fig03}{a}):
For $A_C = \sfrac{\pi}{2}$, the maximum of the six-wave mixing response 
occurs at $\tau_\mathrm{ref} = 2\tau_{12} - \tau_c \approx \SI{65}{\pico\second}$. 
Furthermore, at $A_C = \pi$, 
we can identify the local extremum of the eight-wave-mixing response located at $\tau_\mathrm{ref} = 
2\tau_{12} - 2\tau_c \approx \SI{50}{\pico\second}$. 
Due to the finite duration of the laser pulses, however, the overall temporal envelope of the heterodyning signal has a more complex shape as 
described by Eq.~\eqref{eq: EWM_Ai_2t12}~-~\eqref{eq: 6WM}. We can understand this observation using the approach presented in 
section~\ref{sec: finite_pulses} as a non-trivial interference pattern of multi-wave-mixing contributions that experience a 
different temporal shift caused by the 
finite pulse duration. 
As a special case, when each control pulse area equals $2\pi$, the emission appears 
at $\tau_\mathrm{ref} = 2\tau_{12} - 2 t_p \approx \SI{72}{\pico\second}$, 
which can be understood from our discussion of the finite duration of a single control pulse. 
Here, high-order multi-wave mixing contributions appear advanced by the duration of each control pulse. 
Thus, we are able to trace the transition between perturbative and strong field limit as discussed in section~\ref{sec: finite_pulses}. 

In addition to the temporal modification, we can observe a pulse area dependent 
damping of the signal. The amplitude of the signal's maximum for $A_C = 2\pi$ amounts to roughly 40\% of the photon-echo 
amplitude in the absence of control pulses. 
In a recent study~\cite{grisard_multiple_2022} on the same quantum dot sample, 
we identified phonon-assisted relaxation processes during the action of the laser pulses as the main source for 
the loss of coherence with increasing pulse area.  
In this paper, we set the focus solely on the temporal characteristics of the observed echo signals and 
especially how they are influenced by the pulse area of the control pulses. 
To theoretically reproduce the experimental data we use the non-perturbative 
modeling procedure as applied in section~\ref{sec: finite_pulses}, 
where we approximate the temporal shape of the control pulses as rectangular. Closed expressions 
for the final modeling results are given by Eqs.~(S8)-(S10) in supplementary section 2. 
The temporal cross correlation with the reference pulse is calculated using Eq.~\eqref{eq: CC}. 
To quantify the damping of the signal with increasing pulse area, we took the aforementioned 
damping mechanisms phenomenologically into account by multiplying 
the modeled data with a function $\mathrm{exp}\left(-A_C / A_0\right)$, where the best agreement is found for $A_0 = 2.4 \pi$.
As shown in Fig.~\sref{fig: fig05}{b} we achieve excellent agreement between experimental and modeled data both regarding the temporal 
characteristics of the multi-wave mixing response and the damping of the signal with increasing~$A_C$. 

As discussed in section~\ref{sec: impulsive_limit}, the case $A_C = \sfrac{\pi}{2}$ is in analogy to a Ramsey fringe experiment. 
Figure~\sref{fig: fig05}{c} shows the temporal cross sections of the colormaps in Figs.~\sref{fig: fig05}{a} and (b) 
for $A_C = \sfrac{\pi}{2}$ consisting mainly of three peaks. In the amplitude spectrum, this temporal shape corresponds to spectral 
fringes with a period of roughly $\tau_c^{-1}$ as presented in Fig.~\sref{fig: fig05}{d}. This observation is 
in agreement with our expectations if compared with Figs.~\sref{fig: fig03}{b} and~(c). We thus demonstrate that 
our geometry of optical pulses allows to observe Ramsey fringes on an ensemble of quantum dots. 
Note however that the phase of the signal field $E_S$ is not resolved in our experiments. 
Therefore, we can experimentally access solely the frequency of the Ramsey fringes and the envelope function.
Active stabilization of the optical phase between the two control pulses would allow to gain further control over the 
quantum states of the emitters within the ensemble, which we regard as a prospect for future studies. 
Apart from that, the importance of phase is manifested in the possibility to control the polarization of emission from an ensemble of 
trions as will be demonstrated in the following section.

\section{Polarization sensitive multi-wave mixing}
\begin{figure}
    \centering
    \includegraphics[scale = 1]{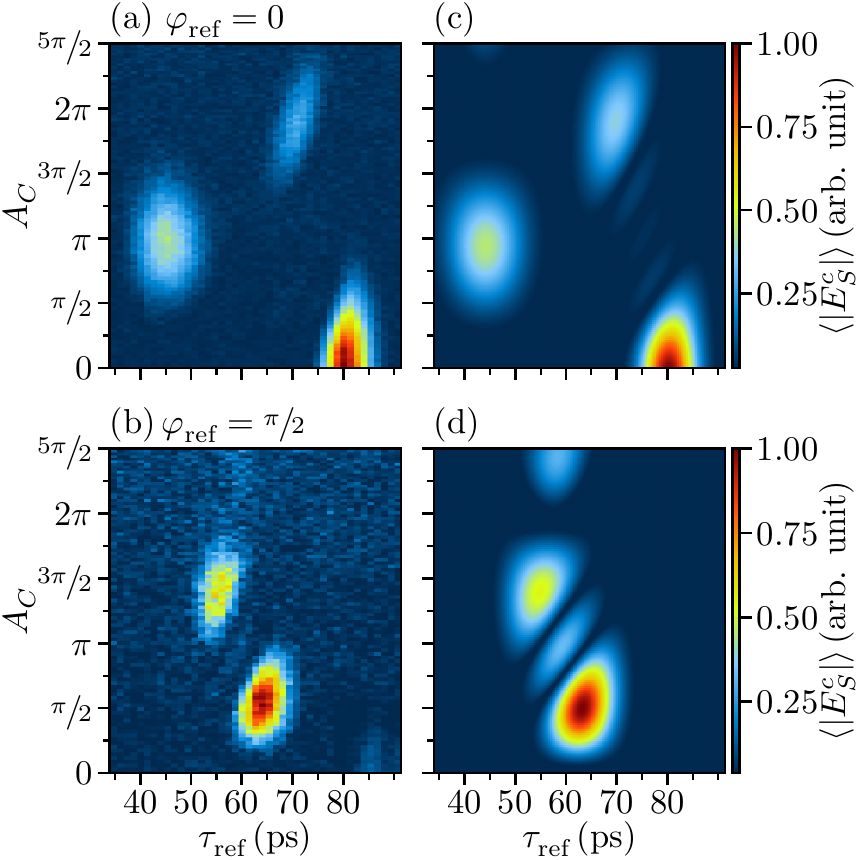}
    \caption{(a)/(b) Heterodyne signal as a function of control pulse area and 
    reference time. In contrast to the measurement in Fig.~5(a), the two control pulses are cross polarized 
    with respect to each other, i.e. polarized horizontally (H) and vertically (V). The signal 
    is detected in H polarization (a) and V polarization (b). (c)/(d) Modeling results corresponding to 
    experimental data in (a) and (b).}
    \label{fig: fig06}
\end{figure}
As a next step, we take into account the spin degree of freedom in the quantum dot sample to expand the 
possibilities of the presented control of multi-wave mixing processes. In what follows, we present 
two possible polarization resolved scenarios, where we show that the selection rules for optical transitions in 
the four-level trion scheme open up new appealing possibilities to address the polarization of the emitted photon 
echoes based on the sensitivity to the relative phase between the two optical control pulses. 

\subsection{Polarization sorting with linearly polarized pulses}

First, we demonstrate that we can use the mutual polarization between the two control pulses to modify the 
polarization of the different multi-wave-mixing processes. 
Therefore, we consider the same pulse arrangement as depicted in Fig.~\sref{fig: fig02}{a}.  
In contrast to the measurements presented so far, the linear polarization angle of the second control pulse with respect 
to the linear polarization of all other pulses is considered as an additional degree of freedom.
To account for polarization in the multi-wave mixing processes from the trion scheme, 
we construct the electric fields of the linear polarized pulses $E_j$ in circular polarization basis
\begin{equation}
    E_j(\varphi_j) = \mathrm{e}^{-i\varphi_j}E^+_j + \mathrm{e}^{+i\varphi_j}E^-_j 
\end{equation}
Without loss of generality, we choose the linear polarizations of exciting, refocussing, and first 
control pulse 
as $\varphi_j = 0$ (horizontally polarized) and introduce $\varphi$ as the relative rotation of the 
polarization of the 
second control pulse. 
We use Eqs.~\eqref{eq: EWM_Ai_2t12}\,-\,\eqref{eq: 6WM} to calculate the eight-wave-mixing 
response at $t = 2\tau_{12} - 2\tau_c$ and 
the six-wave mixing response at $t = 2\tau_{12} - \tau_c$ as a function of $\varphi$ and the polarization $\varphi_\mathrm{ref}$ of the 
reference pulse, which defines the detected polarization component of the signal 
\begin{subequations}
\begin{align}
    E_\mathrm{EWM}(\varphi, \varphi_\mathrm{ref}) &\sim \sin^4\left(\frac{A_C}{2}\right) \cos\left(2\varphi + \varphi_\mathrm{ref} \right) \label{eq: 8WM_phi}\\
    E_\mathrm{SWM}(\varphi, \varphi_\mathrm{ref}) &\sim \sin^2 \left(A_C\right) \cos\left(\varphi + \varphi_\mathrm{ref} \right) \label{eq: 6WM_phi}.
\end{align}
\end{subequations}
Eight- and six-wave mixing thus differ by their dependence on the angle $\varphi$.
In order to experimentally demonstrate this property, 
we choose $\varphi = \sfrac{\pi}{2}$, 
i.e. the control pulses are linearly cross-polarized, which we set as horizontally (H) 
and vertically (V) polarized. 
According to Eqs.~\eqref{eq: 8WM_phi} and~\eqref{eq: 6WM_phi}, 
under this condition the eight-wave mixing response is H-polarized whereas the six-wave mixing signal is
V-polarized. The polarization contrast between both components is thus maximized in the 
configuration $\varphi = \sfrac{\pi}{2}$.
Note that the third main contribution to the overall signal $E_\mathrm{FWM}$, as follows from Eq.~\eqref{eq: EWM_Ai_2t12},
is insensitive to phases of the control pulses and therefore does not dependent on the linear polarization angle $\varphi$. 
The signal $E_\mathrm{FWM}$ is therefore expected to be horizontally polarized as defined by the polarizations of exciting and refocussing pulse.

In Fig.~\ref{fig: fig06} we measured the signal field amplitude 
as a function of $\tau_\mathrm{ref}$ and $A_C$, where we additionally distinguished between 
$\varphi_\mathrm{ref} = 0$ (Fig.~\sref{fig: fig06}{a}, H-detection) and 
$\varphi_\mathrm{ref} = \sfrac{\pi}{2}$ (Fig.~\sref{fig: fig06}{b}, V-detection). 
By comparing the colormaps with the results obtained for co-polarized pulses, Fig.~\sref{fig: fig05}{a}, 
we can observe that we 
achieved to decompose the multi-wave mixing response in H and V polarized components. 
For horizontal detection ($\varphi_\mathrm{ref} = 0$), Fig.~\sref{fig: fig06}{a}, we observe the 2PE at 
$\tau_\mathrm{ref} = 2\tau_{12}$, 
the eight-wave mixing response at $A_C = \pi$, $\tau_\mathrm{ref} = 2\tau_{12} - 2 \tau_{c} = \SI{45}{\pico\second}$, and 
the signal at $A_C = 2\pi$, $\tau_\mathrm{ref} = 2\tau_{12} - 2 t_{p} = \SI{70}{\pico\second}$. 
In contrast, the six-wave-mixing response at 
$A_C = \sfrac{\pi}{2}$, $\tau_\mathrm{ref} - \tau_c \approx \SI{62}{\pico\second}$ 
can be observed solely in vertically polarized detection configuration ($\varphi_\mathrm{ref} = \sfrac{\pi}{2}$), see Fig.~\sref{fig: fig06}{b}. 
Both colormaps~\sref{fig: fig06}{a} and~\sref{fig: fig06}{b} can be accurately reproduced by our 
modeling procedure as discussed above using the optical Bloch equations for the 
trion scheme in Figs.~\sref{fig: fig06}{c} and~\sref{fig: fig06}{d}. 

\subsection{Temporal splitting of the optical response with circularly polarized pulses}
\begin{figure}
    \centering
    \includegraphics[width = \columnwidth]{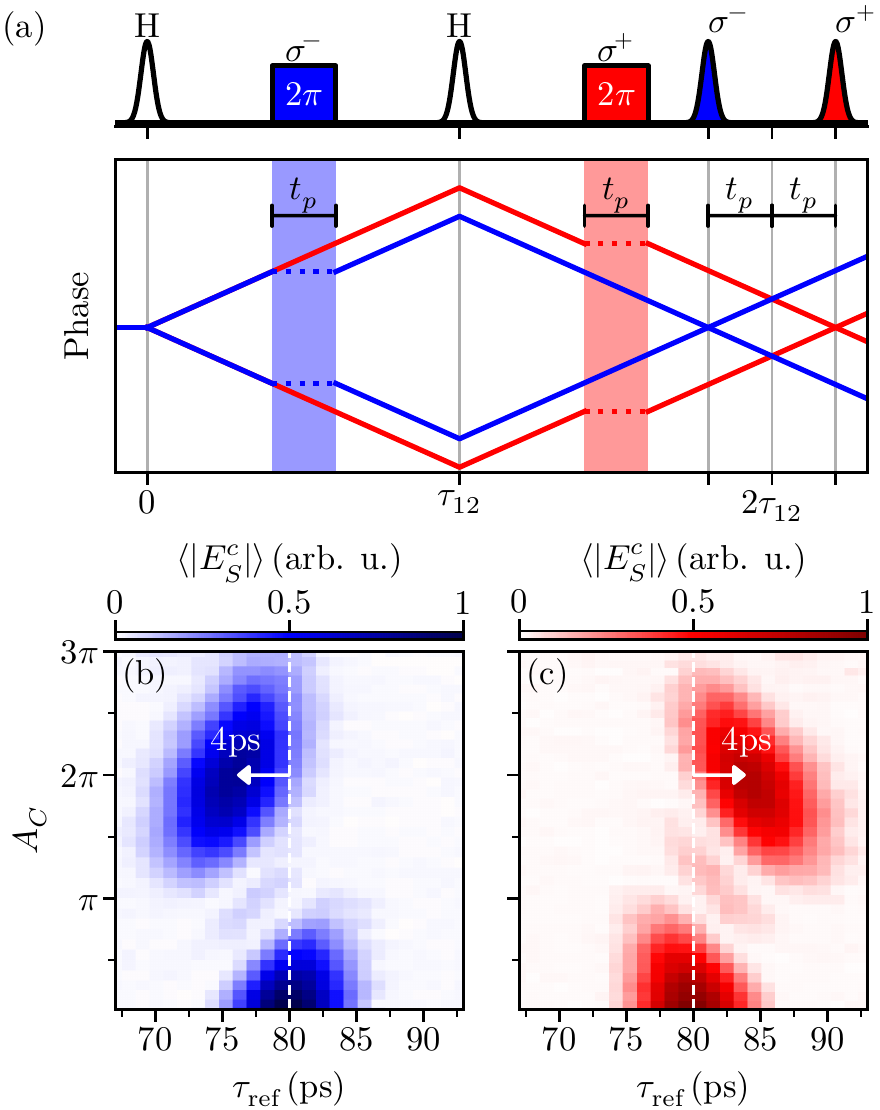}
    \caption{(a) Schematic illustration of the approach 
    for the temporal splitting of linearly polarized photon echo in two cross polarized 
    components.  
    The phase evolution of two oscillators within the $\sigma^-$ ($\sigma^+$) subensemble 
    is shown by blue (red) lines.
    As indicated on top, the first $\sigma^-$ polarized control pulse is temporally located between 
    the exciting and refocussing pulse, whereas the second $\sigma^+$ polarized control pulse acts 
    between the refocussing and photon echo pulse. When the respective areas of the control pulses 
    equal $2\pi$, the phase evolution 
    of the two sub-ensembles ($\sigma^+$ and $\sigma^-$) is effectively frozen for the pulse duration $t_p$. 
    In this 
    way, the coherent response is temporally split in two circularly cross polarized components. Note that all pulses in the experiment 
    have the same temporal shape and duration, the two control pulses are schematically shown as rectangular pulses. 
    (b)/(c) Experimental demonstration of the temporal splitting as discussed in (a).
    For $A_C = 2\pi$, we can observe a shift of the maximum coherent emission by $\mp t_p = \mp\SI{4}{\pico\second}$ for the $\sigma^-$ (b)
    and $\sigma^+$ (c) polarized components as highlighted by the horizontal arrows.}
    \label{fig: fig07}
\end{figure}

As a second example, we elaborate on the possibility to independently modify the coherent emission from 
the two TLS that form the trion scheme using circularly polarized optical pulses.  
In particular, we use circularly polarized control pulses to independently evoke temporal shifts of the 
coherent emission in $\sigma^+$ and $\sigma^-$ polarization. For this purpose, we modify the temporal arrangement 
of the optical pulses as sketched in Fig.~\sref{fig: fig07}{a}. 
Here, the exciting and refocusing pulse  are horizontally polarized, while the control pulses have opposite circular polarizations. The refocusing pulse 
is temporally located in between the two control pulses. This situation corresponds to an experiment with only 
one control pulse for each of the TLS that we considered in section~\ref{sec: finite_pulses}. 
When the area of each control pulse equals $2\pi$, the phase evolution of each sub-ensemble is 
frozen for a time span of $t_p$, which is schematically shown by the phase diagram in Fig.~\sref{fig: fig07}{a}. 
However, for one sub-ensemble, the freezing takes place during the dephasing motion, while for another one 
during the rephasing motion. 
Consequently, the coherent emission is temporally split in two cross polarized components arising 
at $2\tau_{12} \pm t_p$. 

We experimentally demonstrate the effect in Fig.~\ref{fig: fig07} by continuously sweeping the pulse 
area $A_C$ and temporally resolving 
the coherent emission in $\sigma^-$ and $\sigma^+$ detection as shown in Fig.~\sref{fig: fig07}{b} and \sref{fig: fig07}{c}, respectively. 
For a control pulse area of $2\pi$, the $\sigma^-$ component is shifted by $-4$ps, whereas the $\sigma^+$ is shifted by 
$+4$ps relative to the 2PE emission at $2\tau_{12} = \SI{80}{\pico\second}$.
The experiments further allow to observe the 
signal at $A_C = \pi$ that is dominated by a six-wave mixing contribution as discussed 
in section~\ref{sec: finite_pulses} and therefore demonstrates the 
partial lifting 
of degeneracy of wave-mixing orders in the strong field limit by using 
finite pulse durations.  
The overall dependence on $A_C$ and $\tau_\mathrm{ref}$ is well reproduced by the modelled data in 
Fig.~\sref{fig: fig04}{b}. Note that the observed splitting of the photon echo in two 
cross polarized components solely affects the envelope functions of both echo components.

\section{Summary}
We demonstrate coherent optical control over the phase evolution of an ensemble of trions in charged (In,Ga)As quantum dots by 
employing temporally sorted multi-wave-mixing processes that are driven under resonant excitation in the regime of Rabi rotations. 
We have chosen a temporal and angular arrangement of four optical pulses, where two pulses act as 
control knobs of the trion qubits, whereas the other two pulses serve as a photon echo protocol to monitor the phase 
evolution of the inhomogeneously broadened ensemble. Using the energy and momentum conservation, 
we select three photon echo signals that are generated by degenerate four-,~six-, and eight-wave-mixing processes in the lowest non-vanishing 
perturbative order. The temporal delay between the three distinct echo signals is given by the temporal delay between the control pulses. 
Depending on the area of the control pulses $A_C$, the phase evolution of trion ensemble can be modified in several ways. 
Ramsey fringes in the trion ensemble can be observed when 
the area of the control pulses equals $\pi/2$ which is manifested in the time-domain as a coexistence of all three discussed 
echo signals. Next, the phases of trions can be fully inverted by each of the control pulses with area of 
$\pi$ which results in a complete transfer of the macroscopic coherence from the four-wave-mixing echo to 
the eight-wave-mixing echo. 

As a further step, we exploited the spin degree of freedom in a four-level trion scheme to manipulate the polarization state 
of the coherent optical response. These findings clearly demonstrate the importance of the relative optical phase between the 
control pulses, offering the possibility for arbitrary manipulation of the addressed trion states. Further, we 
demonstrate the splitting of the photon echo into two crossed polarized pulses, 
which has potential applications for the transformation between polarization and time bin qubits in integrated semiconductor devices.
We stress that our results are not necessarily limited to the 
optical control of the phase evolution in ensembles of strongly localized excitons or trions as they can be directly adopted 
to control the phase evolution and the corresponding optical response of a single quantum emitter. Our demonstrations 
push forward the realization of arbitrary pulse sequences, as widely used in NMR, for quantum memory protocols or information 
processing in ensembles of quantum dots by optical methods. We furthermore suggest new possibilities for coherent control that 
are not subject to NMR, such as polarization and wave vector selectivity.

We consider the importance of the frequency detuning of the driving optical field during action of the control pulses. 
Here, using a perturbative multi-wave-mixing expansion we gain insight into how the coherent emission is modified 
when the inhomogeneous broadening of the excited ensemble and the width of the laser spectrum are comparable. Surprisingly, 
we reveal that different orders of multi-wave mixing lead to an additional temporal sorting of the optical response, 
which can be interpreted in terms of additional modifications of the phase evolution. 
Lifting of the temporal degeneracy between different wave-mixing 
processes allows to trace the transition from perturbative to strong field regime with Rabi rotations and opens 
up new possibilities for optical investigations of complex energy structures in unexplored material systems, e.g. excitons 
in heterostructures based on transition metal dichalcogenides~\cite{hao_direct_2016, katsch_optical_2020} or 
lead-halide perovskites~\cite{raino_superfluorescence_2018, liu_multidimensional_2021}.

\section*{Data availability statement}
The data that support the findings of this study are available from the corresponding
author upon reasonable request.

\begin{acknowledgments}
We acknowledge financial support from the Deutsche
Forschungsgemeinschaft (DFG) through the Collaborative
Research Center TRR 142/3 (Grant No. 231447078, Project
No. A02) and from the Bundesministerium für Bildung und
Forschung (BMBF) within the project “QR.X” (Project Nos.
16KISQ010 and 16KISQ011). 
We are grateful to D.~Suter and I.~A.~Yugova for useful discussions.
\end{acknowledgments}

\nocite{*}

\clearpage

\ifarXiv
    

\section*{Supplementary material}

# Supplementary information:
# Temporal sorting of optical multi-wave-mixing processes in semiconductor quantum dots

S. Grisard[1], A. V. Trifonov[1], H. Rose[2], R. Reichhardt[1], M. Reichelt[2],
C. Schneider[3, 4], M. Kamp[3], S. Höfling[3], M. Bayer[1], T. Meier[2], and
I. A. Akimov[1]

[1]Experimentelle Physik 2, Technische Universität Dortmund, 44221
Dortmund, Germany
[2]Paderborn University, Department of Physics & Institute for Photonic
Quantum Systems (PhoQS), 33098 Paderborn, Germany
[3]Technische Physik, Universität Würzburg, 97074 Würzburg, Germany
[4]Institute of Physics, University of Oldenburg, 26129 Oldenburg,
Germany

# 1 Characterization of optical control pulses

Figure S1 shows the measured cross-correlation between the reference pulse and the two control pulses. The intensity signal on the photodetector results from interference between the reference pulse $E_{\text{ref}}$ and control pulse fields $E_{\text{C1}}$, $E_{\text{C2}}$

$$I_{\text{CC}}(\tau_{\text{ref}}) \propto \int dt (E_{\text{C1}}(t) + E_{\text{C2}}(t)) E_{\text{ref}}^*(t - \tau_{\text{ref}}). \tag{S1}$$

As shown by the black solid line, the signal is well described by the sum of two Gaussians with FWHM of 5.6 ps temporally separated by $\tau_c = 15.3$ ps. Assuming that all pulses have the same duration, we can extract a pulse duration of $5.6\,\text{ps}/\sqrt{2} \approx 4.0\,\text{ps}$. The unconvoluted pulse envelopes are shown by the green dashed line in Fig. S1.

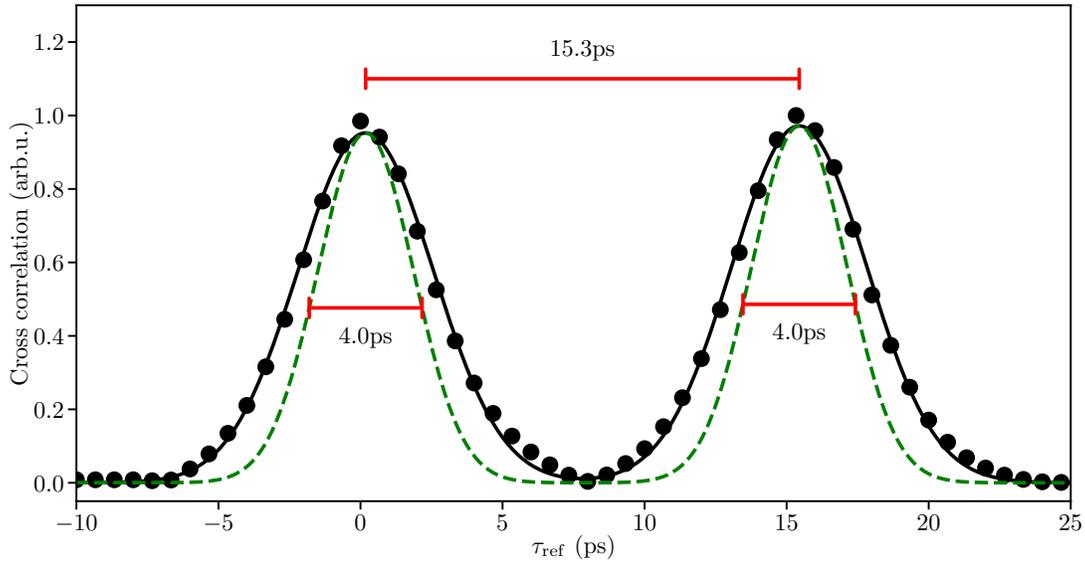

**Figure S1:** Cross-correlation between the reference pulse and the two control pulses. Black dots show the experimental data, black line is a fit by two Gaussians, green lines are the deconvoluted pulse envelopes.

# 2 Action of rectangular pulses on two-level system

This section provides the equations used for the modeling results presented in Figs. 4(b)/(c), 5(b), and 6(c)/(d) of the main text. Here, the effect of optical control pulses acting between the two pulses of a regular photon echo sequence is discussed as a function of time and pulse area of the control pulses. Importantly, the duration of the control pulse is comparable to the dephasing time $T_2^*$ of an ensemble of emitters, which are modeled as two-level systems. In this case, the dephasing of the ensemble during the pulse action is not negligible. For simplicity, the temporal shape of the control pulse is assumed to be rectangular, for which analytical solutions of the optical Bloch equations are well known (Rabi solution [43]), the relevant equations are presented in the following.

We describe the quantum state of a two level system with states $|0\rangle$ and $|1\rangle$, using the corresponding density matrix $\rho_{ij} = |i\rangle\langle j|$. The action of a rectangular shaped pulse can be

written as a matrix operation on the elements of the density matrix that transforms the density matrix elements before the action of the pulse (index b) into the density matrix elements after the action of the pulse (index a)

$$
\begin{pmatrix} \rho_{00} \\ \rho_{11} \\ \rho_{01} \\ \rho_{10} \end{pmatrix}_a = \mathbf{M} \begin{pmatrix} \rho_{00} \\ \rho_{11} \\ \rho_{01} \\ \rho_{10} \end{pmatrix}_b , \tag{S2}
$$

where the matrix $\mathbf{M}$ is given by

$$
\mathbf{M} = \begin{pmatrix}
1 - \frac{\Omega_R^2(1-c)}{2\Omega^2} & \frac{\Omega_R^2(1-c)}{2\Omega^2} & \left( -\frac{\nu\Omega_R(1-c)}{2\Omega^2} + i\frac{\Omega_R s}{2\Omega} \right) e^{-i\phi} & \left( -\frac{\nu\Omega_R(1-c)}{2\Omega^2} - i\frac{\Omega_R s}{2\Omega} \right) e^{+i\phi} \\
\frac{\Omega_R^2(1-c)}{2\Omega^2} & 1 - \frac{\Omega_R^2(1-c)}{2\Omega^2} & \left( \frac{\nu\Omega_R(1-c)}{2\Omega^2} - i\frac{\Omega_R s}{2\Omega} \right) e^{-i\phi} & \left( \frac{\nu\Omega_R(1-c)}{2\Omega^2} + i\frac{\Omega_R s}{2\Omega} \right) e^{+i\phi} \\
\left( -\frac{\nu\Omega_R(1-c)}{2\Omega^2} + i\frac{\Omega_R s}{2\Omega} \right) e^{+i\phi} & \left( \frac{\nu\Omega_R(1-c)}{2\Omega^2} - i\frac{\Omega_R s}{2\Omega} \right) e^{+i\phi} & c + \frac{\Omega_R^2(1-c)}{2\Omega^2} + i\frac{\nu s}{\Omega} & \frac{\Omega_R^2(1-c)}{2\Omega^2} e^{2i\phi} \\
\left( -\frac{\nu\Omega_R(1-c)}{2\Omega^2} - i\frac{\Omega_R s}{2\Omega} \right) e^{-i\phi} & \left( \frac{\nu\Omega_R(1-c)}{2\Omega^2} + i\frac{\Omega_R s}{2\Omega} \right) e^{-i\phi} & \frac{\Omega_R^2(1-c)}{2\Omega^2} e^{-2i\phi} & c + \frac{\Omega_R^2(1-c)}{2\Omega^2} - i\frac{\nu s}{\Omega}
\end{pmatrix} \tag{S3}
$$

With the detuning between laser and transition frequency $\nu$, Rabi frequency $\Omega_R$, generalized Rabi frequency $\Omega = \sqrt{\Omega_R^2 + \nu^2}$, pulse duration $t_p$, $c = \cos\Omega t_p$, $s = \sin\Omega t_p$, and the optical phase of the pulse $\phi = \mathbf{k} \cdot \mathbf{r} + \varphi$ including the spatial phase $\mathbf{k} \cdot \mathbf{r}$ and an additional optical phase $\varphi$. Multiple applications of Eq. (S2) can be used to model an arbitrary sequence of optical pulses. The free evolution of the density matrix elements during a temporal gap $\tau$ between two pulses is obtained by setting $t_p = \tau$ and $\Omega_R = \phi = 0$ in Eq. (S3). The final optical response is proportional to the off-diagonal element of the density matrix $\rho_{01}$, from which the contributions including the desired phase matching condition, in our case $2\mathbf{k}_2 - \mathbf{k}_1$, can be extracted. The measured macroscopic response $E_S$ from an ensemble of two-level systems is proportional to the integral over all detunings $\nu$

$$
E_S \propto \int \rho_{01} G(\nu) \mathrm{d}\nu, \tag{S4}
$$

where $G$ denotes the distribution of detunings. Here, we assume a Gaussian distribution with FWHM given by the dephasing time $T_2^*$

$$
G(\nu) = \frac{1}{\sqrt{2\pi}\sigma} e^{-\frac{\nu^2}{2\sigma^2}}, \quad \sigma = \frac{T_2^*}{\sqrt{8\ln 2}}. \tag{S5}
$$

As discussed in the main text, we consider the situation where, the spectrum of the laser pulses is significantly more narrow than the spectrum of the inhomogeneously broadened ensemble. Therefore, the laser spectrum effectively selects a subensemble where $T_2^*$ equals the pulse duration $t_p$.

Following the described approach, we can provide closed expressions for the modeling results shown in the main text:

- **Fig. 4(b)**: Dependence of the signal field emitted in direction $2k_2 - k_1$ as a function of the area $A_C$ of a single control pulse and real time $\tau$:

$$
E_S \propto \int_{-\infty}^{\infty} \mathrm{d}\nu\, G(\nu) \left\{ \frac{\nu}{\Omega_C} \sin\left(\Omega_C t_p\right) + i \left[ \frac{\Omega_{R,C}^2}{\Omega_C^2} \sin^2\left(\frac{\Omega_C t_p}{2}\right) + \cos\left(\Omega_C t_p\right) \right] \right\} e^{i\nu(\tau_{\mathrm{ref}} - 2\tau_{12} + t_p)}, \tag{S6}
$$

where $\Omega_C = \sqrt{\Omega_{R,C}^2 + \nu^2}$, $\Omega_{R,C} = A_C/t_p$. The integral over detunings $\nu$ is carried out numerically using the trapezoidal rule. The cross sections shown in Fig. 4(c) can be obtained by inserting the corresponding values of $A_C$ in Eq. (S6).

- **Fig. 5(b)**: The signal field in an experiment with two control pulses of equal area $A_C$ can be written as a sum of three contributions

$$E_S = E_{\text{FWM}} + E_{\text{SWM}} + E_{\text{EWM}} \tag{S7}$$

where

$$
\begin{aligned}
E_{\text{FWM}} \propto \int_{-\infty}^{\infty} \mathrm{d}\nu\, G(\nu) \Bigg\{ &\frac{\Omega_{R,C}^2 \nu}{\Omega_C^3} \left[ \sin(\Omega_C t_p)\cos(\Omega_C t_p) - \sin(\Omega_C t_p) \right] \\
&- \frac{\nu}{\Omega_C}\sin(\Omega_C t_p)\cos(\Omega_C t_p) - i\frac{\Omega_{R,C}^4}{2\Omega_C^4}\sin^4\left(\frac{\Omega_C t_p}{2}\right) \\
&+ i\frac{\nu^2}{2\Omega_C^2}\sin^2(\Omega_C t_p) + i\frac{\Omega_{R,C}^2}{2\Omega_C^2}\cos(\Omega_C t_p)[\cos(\Omega_C t_p) - 1] \\
&- i\frac{1}{2}\cos^2(\Omega_C t_p) \Bigg\} \mathrm{e}^{i\nu(\tau_{\text{ref}} - 2\tau_{12} + 2t_p)}
\end{aligned} \tag{S8}
$$

$$
\begin{aligned}
E_{\text{SWM}} \propto \int_{-\infty}^{\infty} \mathrm{d}\nu\, G(\nu) \Bigg\{ &\frac{\Omega_{R,C}^2 \nu}{2\Omega_C^3}[\sin(\Omega_C t_p) - \sin(\Omega_C t_p)\cos(\Omega_C t_p)] \\
&+ i\left[ \frac{\Omega_{R,C}^2}{4\Omega_C^2}\sin^2(\Omega_C t_p) - \frac{\Omega_{R,C}^2 \nu^2}{\Omega_C^4}\sin^4\left(\frac{\Omega_C t_p}{2}\right) \right] \Bigg\} \times \\
&\times \mathrm{e}^{i\nu(\tau_{\text{ref}} - 2\tau_{12} + \tau_c + t_p) - i\varphi_c}
\end{aligned} \tag{S9}
$$

$$
E_{\text{EWM}} \propto \int_{-\infty}^{\infty} \mathrm{d}\nu\, G(\nu) \left\{ -i\frac{\Omega_{R,C}^4}{2\Omega_C^4}\sin^4\left(\frac{\Omega_C t_p}{2}\right) \right\} \mathrm{e}^{i\nu(\tau_{\text{ref}} - 2\tau_{12} + 2\tau_c) - i2\varphi_c} \tag{S10}
$$

Here, $\tau_c$ denotes the temporal gap between the control pulses and $\varphi_c$ the relative optical phase between the control pulses. To directly compare with the experimental data, that is obtained by bringing the signal field to interference with a reference pulse, we calculate the temporal cross correlation of the modelled data (S7) with a Gaussian reference pulse of FWHM $t_p = 4ps$.

- **Fig. 6(c)/(d)**: Equation (S7) is also used for the polarization sensitive measurements on the trion ensemble presented in section IV B. As discussed in the main text, the trion scheme consists of two independent two-level systems that can be independently addressed by $\sigma^+$ and $\sigma^-$ polarized light. Thus, an arbitrary linear polarization angle $\varphi$ between the control pulses translates into a relative phase between $\sigma^+$ and $\sigma^-$ components of the final radiation, i.e. a particular polarization of the emitted light. The polarization selection rules of the three different terms $E_{\text{FWM}}$, $E_{\text{SWM}}$, and $E_{\text{EWM}}$ that we discussed in the main text in the impulsive limit keep their validity also when taking into account the finite pulse duration as can be seen from the dependence on the relative phase $\varphi_c$ in Equations (S8)–(S10).

Finally, we note that the analytical results presented in section IIA, Eqs. (3)-(6), can be obtained in an analogous way. Here, however, we neglect the finite duration of the pulses. Formally, this can be achieved by setting $\nu = 0$ in Eq. (S3) for the pulse action and replacing $\Omega_R t_p$ by the pulse area $A$ of the involved delta pulses defined as $A = \int \frac{\mu}{\hbar} E_0 \nu(t - \tau)dt$. The

matrix $\mathbf{M}$ simplifies to

$$\mathbf{M} = \frac{1}{2} \begin{pmatrix} 1 + \cos(A) & 1 - \cos(A) & i\sin(A)\mathrm{e}^{-i\phi} & -i\sin(A)\mathrm{e}^{+i\phi} \\ 1 - \cos(A) & 1 + \cos(A) & -i\sin(A)\mathrm{e}^{-i\phi} & i\sin(A)\mathrm{e}^{+i\phi} \\ i\sin(A)\mathrm{e}^{+i\phi} & -i\sin(A)\mathrm{e}^{+i\phi} & 1 + \cos(A) & (1 - \cos(A))\mathrm{e}^{2i\phi} \\ -i\sin(A)\mathrm{e}^{-i\phi} & i\sin(A)\mathrm{e}^{-i\phi} & (1 - \cos(A))\mathrm{e}^{-2i\phi} & 1 + \cos(A). \end{pmatrix} \quad \text{(S11)}$$

# 3 Multi-wave-mixing expansion

This section presents in detail the perturbative multi-wave-mixing approach that is used in section IIB of the main text to gain insight into the effect of finite pulse duration in the control pulse experiments. The description of non-linear optical phenomena by means of multi-wave-mixing processes relies on the expansion of the nonlinear polarization into orders with respect to the amplitudes of the optical fields that interact with the system. Particular multi-wave-mixing processes that can be distinguished for example making use of the phase matching condition, are categorized using double sided Feynman diagrams. Here, we want to give a brief derivation of the diagrams and how they can be used to keep track of all terms that need to be taken into account to calculate a particular nonlinear polarization. Our approach is based on references [48, 49, 54].

The microscopic polarization $P$ of a quantum system (in our case a two-level system) at observation time $t$ is given by the expectation value of the dipole operator $\hat{\mu}(t)$

$$P = \langle \hat{\mu}(t) \rangle = \mathrm{Tr}(\hat{\mu}(t)\hat{\rho}(t)), \quad \text{(S12)}$$

where the time dependent density matrix operator $\hat{\rho}(t) = |i\rangle\langle j|$ results from the Liouville-von Neumann equation (formulated in interaction picture)

$$\frac{d\hat{\rho}(t)}{dt} = \frac{i}{\hbar}[\hat{\rho}(t), \hat{V}(t)]. \quad \text{(S13)}$$

With the interaction operator $\hat{V}$ resulting from the interaction of the system with $N$ optical fields

$$\hat{V}(t) = \sum_{m=0}^{N} \hat{V}_m = -\frac{\hat{\mu}(t)}{2} \sum_{m=0}^{N} \left( E_m(t)\mathrm{e}^{-i\mathbf{k}_m \cdot \mathbf{r}} + E_m^*(t)\mathrm{e}^{i\mathbf{k}_m \cdot \mathbf{r}} \right). \quad \text{(S14)}$$

Here, $E_m(t)$ denote the electric field envelopes and $\mathbf{k}_m$ the corresponding wave vectors. The Hamiltonian of the unperturbed system $\hat{H}_0$ in rotating wave frame is given by

$$\hat{H}_0 = \hbar \begin{pmatrix} -\frac{\nu}{2} & 0 \\ 0 & \frac{\nu}{2} \end{pmatrix}, \quad \text{(S15)}$$

with detuning $\nu$ between laser and transition frequency.

A perturbative expansion of $P$ relies on the expansion of the density matrix $\hat{\rho}$ in orders $\hat{\rho}^{(i)}$ with respect to powers of the interacting electric field amplitudes

$$P = \mathrm{Tr}(\hat{\mu}(t)\hat{\rho}^{(0)}(t)) + \mathrm{Tr}(\hat{\mu}(t)\hat{\rho}^{(1)}(t)) + \cdots + \mathrm{Tr}(\hat{\mu}(t)\hat{\rho}^{(n)}(t)) + \dots. \quad \text{(S16)}$$

As follows from Equation (S13), the $n$th order can be written as

$$\hat{\rho}^{(n)}(t) = \left(\frac{1}{i\hbar}\right)^n \int_{-\infty}^{t} dt_n \int_{-\infty}^{t_n} dt_{n-1} \cdots \int_{-\infty}^{t_2} dt_1 [\hat{V}(t_n), [\hat{V}(t_{n-1}), [\cdots [\hat{V}(t_1), \hat{\rho}(0)] \cdots]]], \quad \text{(S17)}$$

Using Equations (S14)–(S17), the $n$th order polarization is consequently given by

$$P^{(n)} = \int_{-\infty}^{t} dt_n \int_{-\infty}^{t_n} dt_{n-1} \cdots \int_{-\infty}^{t_2} dt_1 R^{(n)}(t, t_n, \dots, t_1) \mathcal{E}_n(t_n) \cdots \mathcal{E}_1(t_1) \tag{S18}$$

with $\mathcal{E}_m = E_m e^{-i\mathbf{k}_m \cdot \mathbf{r}} + E_m^* e^{i\mathbf{k}_m \cdot \mathbf{r}}$ and the response function

$$R^{(n)} = \left(\frac{i}{2\hbar}\right)^n \operatorname{Tr}\left([\dots [\hat{\mu}(t), \hat{\mu}(\tau_n)], \dots, \hat{\mu}(\tau_1)] \hat{\rho}(0)\right). \tag{S19}$$

Note that Equation (S18) assumes that the $N$ electric fields are temporally ordered. In this way, the number of summands contributing to $P^{(n)}$ is reduced from $2^{2n} N^n$ to $2^{2n}$ remaining from combinations of $E_m$ and $E_m^*$ in (S18) and elements of the commutator products in (S19). Additionally, the number can be strongly reduced by a restriction to a particular phasematching condition.

A well established tool to visualize the different summands that contribute to the nonlinear polarization are double sided Feynman diagrams. In the following, we briefly discuss how these diagrams are motivated. For this purpose, we consider the special case of the four-wave-mixing response $P^{(3)} \propto E_1 \left(E_2^*\right)^2$ of a two-level system under action of two laser pulses. Further, we assume delta-like pulses $E_1(t) \propto \delta(t)$ and $E_2(t) \propto \delta(t - \tau_{12})$, $\tau_{12} > 0$. The third-order polarization in this case reads as

$$P^{(3)} = \left(\frac{i}{2\hbar}\right)^3 \operatorname{Tr}\left([[[\hat{\mu}(t), \hat{\mu}(\tau_{12})], \hat{\mu}(\tau_{12})], \hat{\mu}(0)] \hat{\rho}(0)\right) E_1 \left(E_2^*\right)^2 e^{i(2\mathbf{k}_2 - \mathbf{k}_1) \cdot \mathbf{r}} \tag{S20}$$

which includes for example the summand:

$$P_1^{(3)} = \left(\frac{i}{2\hbar}\right)^3 \operatorname{Tr}\left(\hat{\mu}(0) \hat{\mu}(\tau_{12}) \hat{\mu}(t) \hat{\mu}(\tau_{12}) \hat{\rho}(0)\right) E_1 \left(E_2^*\right)^2 e^{i(2\mathbf{k}_2 - \mathbf{k}_1) \cdot \mathbf{r}} \tag{S21}$$

We can explicitly write the time-expanded dipole operators using $\hat{\mu}(t) = \mathbf{U}^\dagger(t) \hat{\mu} \mathbf{U}(t)$ with $\mathbf{U} = \exp\left(i \frac{\mathbf{H}_0}{\hbar} t\right)$. Further, we can make use of the invariance of the trace operation under cyclic permutations to rewrite $P_1^{(3)}$ as

$$P_1^{(3)} = \left(\frac{i}{2\hbar}\right)^3 \operatorname{Tr}\left\{\hat{\mu} \left[\mathbf{U}(t - \tau_{12}) \left[\hat{\mu} \left[\mathbf{U}(\tau_{12}) \left[\hat{\rho}(0) \hat{\mu}\right] \mathbf{U}^\dagger(\tau_{12})\right] \hat{\mu}\right] \mathbf{U}^\dagger(t - \tau_{12})\right]\right\} E_1 \left(E_2^*\right)^2 e^{i(2\mathbf{k}_2 - \mathbf{k}_1) \cdot \mathbf{r}}. \tag{S22}$$

In this form, we can understand the underlying process as follows. First, the dipole operator $\hat{\mu}$ acts on the right side of the initial density matrix (interaction with $E_1$; blue brackets). The resulting density matrix then freely evolves during the temporal gap between first and second pulse $\tau_{12}$ (green brackets). Subsequently, the dipole operator $\mu$ acts from left and right side on the density matrix (twice interaction with $E_2^*$; red brackets). The density matrix then again freely evolves between the second pulse and observation time $t$ for a duration of $t - \tau_{12}$ (purple brackets). Finally, the density matrix is multiplied again by $\hat{\mu}$ from the left side (final emission). For a two-level system, where initially only the ground state $|0\rangle$ it populated, such process can be visualized by the following Feynman diagram

$$\tag{S23}$$

Here, time runs from top to bottom, and arrows indicate interactions with the fields (mathematically, multiplications from right or left side with the dipole operator). The direction of the arrows with label $k_j$ indicate whether $E_j e^{-i\mathbf{k}_j\cdot\mathbf{r}}$ or $E_j^* e^{i\mathbf{k}_j\cdot\mathbf{r}}$ contribute. In this way, the phase matching condition for the final emission can be directly read off. We can use the diagrams like shown above as a short and clear form of writing summands like as given in Eq. (S22). Every diagram of a two level system can be constructed from eight unique elements that are displayed in table S1. Each element corresponds to a certain factor to the kernel of the multi-time integral appearing in the nonlinear polarization (S18). When the finite duration of the optical pulses is neglected, the integral kernel directly corresponds to the polarization due to the delta functions in the temporal envelopes of the electric fields.

**Table S1:** Unique elements of Feynman diagrams for a two-level system. The corresponding kernel for calculation of the non-linear polarization can be constructed by multiplying the factors given in the table. The dipole moment and $\hbar$ are set to 1.

| Element | Factor | Element | Factor |
|---|---|---|---|
| $k \rightarrow \begin{vmatrix}|0\rangle\langle0| \\ |1\rangle\langle1|\end{vmatrix}\begin{matrix}---t_{m-1}\\---t_m\\---t_{m+1}\end{matrix}$ | $\frac{i}{2}E_m^*(t_m)e^{i\mathbf{k}\cdot\mathbf{r}+i\nu(t_{m+1}-t_m)}$ | $\begin{matrix}t_{m-1}---\\t_m---\\t_{m+1}---\end{matrix}\begin{vmatrix}|0\rangle\langle0| \\ |0\rangle\langle1|\end{vmatrix}\leftarrow k$ | $-\frac{i}{2}E_m(t_m)e^{-i\mathbf{k}\cdot\mathbf{r}-i\nu(t_{m+1}-t_m)}$ |
| $k \rightarrow \begin{vmatrix}|0\rangle\langle1| \\ |1\rangle\langle1|\end{vmatrix}\begin{matrix}---t_{m-1}\\---t_m\\---t_{m+1}\end{matrix}$ | $\frac{i}{2}E_m^*(t_m)e^{i\mathbf{k}\cdot\mathbf{r}}$ | $\begin{matrix}t_{m-1}---\\t_m---\\t_{m+1}---\end{matrix}\begin{vmatrix}|1\rangle\langle0| \\ |1\rangle\langle1|\end{vmatrix}\leftarrow k$ | $-\frac{i}{2}E_m(t_m)e^{-i\mathbf{k}\cdot\mathbf{r}}$ |
| $k \leftarrow \begin{vmatrix}|1\rangle\langle0| \\ |0\rangle\langle0|\end{vmatrix}\begin{matrix}---t_{m-1}\\---t_m\\---t_{m+1}\end{matrix}$ | $\frac{i}{2}E_m(t_m)e^{-i\mathbf{k}\cdot\mathbf{r}}$ | $\begin{matrix}t_{m-1}---\\t_m---\\t_{m+1}---\end{matrix}\begin{vmatrix}|0\rangle\langle1| \\ |0\rangle\langle0|\end{vmatrix}\rightarrow k$ | $-\frac{i}{2}E_m^*(t_m)e^{i\mathbf{k}\cdot\mathbf{r}}$ |
| $k \leftarrow \begin{vmatrix}|1\rangle\langle1| \\ |0\rangle\langle1|\end{vmatrix}\begin{matrix}---t_{m-1}\\---t_m\\---t_{m+1}\end{matrix}$ | $\frac{i}{2}E_m(t_m)e^{-i\mathbf{k}\cdot\mathbf{r}-i\nu(t_{m+1}-t_m)}$ | $\begin{matrix}t_{m-1}---\\t_m---\\t_{m+1}---\end{matrix}\begin{vmatrix}|1\rangle\langle1| \\ |1\rangle\langle0|\end{vmatrix}\rightarrow k$ | $-\frac{i}{2}E_m^*(t_m)e^{i\mathbf{k}\cdot\mathbf{r}+i\nu(t_{m+1}-t_m)}$ |

In the main text, we considered the action of a single control pulse applied temporally between the two pulses $E_1$ and $E_2$ of a photon echo sequence. Here, we analyzed how the echo response is temporally modified when the control pulse is considered in different orders with respect to the control pulse field $E_C$. We only capture signals emitted in the direction $2\mathbf{k}_2 - \mathbf{k}_1$. Therefore, we take into account only those nonlinear terms, that involve the same number of interactions with $E_C$ and $E_C^*$, such as six-wave mixing $P^{(5)} \propto E_1(E_2^*)^2 E_C E_C^*$. For the six-wave-mixing process, there are four unique Feynman diagrams that fulfill the time ordering and phase-matching condition. We show these diagrams in table S2. We consider the first and second pulse as delta pulses, centered at $t=0$ and $t=\tau_{12}$, respectively. All diagrams give the same contribution to the nonlinear polarization as a function of detuning $\nu$, that is

$$P^{(5)}(t,\nu) \propto E_1(E_2^*)^2 \int_{-\infty}^{\tau_{12}} dt_{C2} \int_{-\infty}^{t_{C2}} dt_{C1} E_C(t_{C2})E_C^*(t_{C1})e^{i\nu(t-2\tau_{12})}e^{i\nu(t_{C2}-t_{C1})} \tag{S24}$$

where we labeled the integration variables corresponding to the control pulse field with $t_{Cj}$. Here, we can already see, that the control pulse has a direct impact on the phase evolution of the ensemble, as the usual phase factor $e^{i\nu(t-2\tau_{12})}$ that gives rise to a photon echo centered at $t=2\tau_{12}$ is modified by the factor $e^{i\nu(t_{C2}-t_{C1})}$. This leads to a shifting of the maximum echo response as we highlighted in the main text. For the particular temporal shape of the control pulse field $E_C(t)$ we assumed for simplicity a rectangular shape with a width of $t_p$. This step significantly simplifies the calculation effort, but may lead to deviations from the experiment, where Gaussian pulses were used. To finally obtain the overall response of the

inhomogeneous ensemble, we numerically average the nonlinear polarization over a Gaussian distribution of detuning as described in section 2 of the supplement. In this way we obtain the echo response of the ensemble in second order perturbation with respect to the control pulse field, which is shown in Fig. 4(d) of the main text (red line).

**Table S2:** Unique Feynman diagrams for the six-wave-mixing process in a single control pulse experiment fulfilling the phase matching condition $2k_2 - k_1$ and time ordering of $E_1$, $E_C$, and $E_2$.

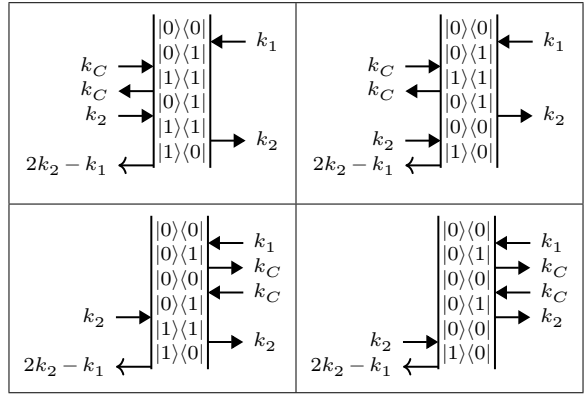

In the same manner as for six-wave mixing, we obtained the echo pulses for higher order interactions with the control pulse field. The next orders are eight- and ten-wave-mixing processes including four and six interactions with the control pulse fields, respectively. For these two orders, there are 16 and 64 Feynman diagrams that we explicitly show in Tables S3 and S4, respectively. Note that for higher orders than six-wave mixing, not all Feynman diagrams give necessarily the same response function. Compare for example in Table S3 the first and second diagram in the top row. Here, the phase evolution between second and third interaction with the control pulse field is conjugated. Since the number of Feynman diagrams increases exponentially with increasing order, we developed a python-program that obtains all possible Feynman diagrams given a specific perturbative order, phase matching condition, and temporal ordering of pulses. The source code is available on reasonable request.

**Table S3:** Unique Feynman diagrams for the eight-wave mixing process in a single control pulse experiment fulfilling the phase matching condition $2k_2 - k_1$ and time ordering of $E_1$, $E_C$, and $E_2$.

**Table S4:** Unique Feynman diagrams for the ten-wave mixing process in a single control pulse experiment fulfilling the phase matching condition $2k_2 - k_1$ and time ordering of $E_1$, $E_C$, and $E_2$.


\fi
\end{document}